\documentclass[final,12pt]{elsarticle}

\usepackage{multicol} 
\usepackage{multirow} 
\usepackage{rotating} 
\usepackage{amsmath} 
\usepackage{stmaryrd} 
\usepackage{color} 
\definecolor{darkred}{rgb}{0.85, 0.0, 0.0}
\definecolor{darkgreen}{rgb}{0.0, 0.6, 0.0}
\usepackage{float} 
\usepackage{booktabs} 
\usepackage{colortbl} 
\usepackage{url}
\usepackage[margin=3cm]{geometry}
\usepackage{tabularx}  
\usepackage{longtable}
\usepackage{ltcaption}

\usepackage[figure]{algorithm2e} 
\usepackage{threeparttable} 
\usepackage{colortbl} 

\biboptions{square}

\journal{arXiv.org}

\begin{document}


\begin{frontmatter}



\title{Improving type information inferred by decompilers \\ with supervised machine learning}

\author[affil1]{Javier Escalada}
\ead{escaladajavier@uniovi.es}

\author[affil2]{Ted Scully}
\ead{ted.scully@cit.ie}
\ead[url]{http://cs.cit.ie/research-staff.ted-scully.biography}

\author[affil1,affil2]{Francisco Ortin\corref{cor1}}
\ead{ortin@uniovi.es}
\ead[url]{http://www.reflection.uniovi.es/ortin}
\cortext[cor1]{Corresponding author}

\address[affil1]{University of Oviedo, Computer Science Department, \\c/Federico Garcia Lorca 18, 33007, Oviedo, Spain\\ $ $ }
\address[affil2]{Cork Institute of Technology, Computer Science Department, \\ Rossa Avenue, Bishopstown, Cork, Ireland}



\begin{abstract}

In software reverse engineering, decompilation is the process of recovering source code from binary files. Decompilers are used when it is necessary to understand or analyze software for which the source code is not available. Although existing decompilers commonly obtain source code with the same behavior as the binaries, that source code is usually hard to interpret and certainly differs from the original code written by the programmer. Massive codebases could be used to build supervised machine learning models aimed at improving existing decompilers. In this article, we build different classification models capable of inferring the high-level type returned by functions, with significantly higher accuracy than existing decompilers. We automatically instrument C source code to allow the association of binary patterns with their corresponding high-level constructs. A dataset is created with a collection of real open-source applications plus a huge number of synthetic programs. Our system is able to predict function return types with a 79.1\% F\textsubscript{1}-measure, whereas the best decompiler obtains a 30\% F\textsubscript{1}-measure. Moreover, we document the binary patterns used by our classifier to allow their addition in the implementation of existing decompilers.

\end{abstract}


\end{frontmatter}


\section{Introduction}
\label{section:introduction}

A decompiler is a tool that receives binary code as input and generates high-level code with the same semantics as the input. Although the decompiled source code can be recompiled to produce the original binary code, the high-level source code is not commonly the one originally written by the programmer. In fact, the source code is usually much less readable than the original one~\cite{Horspool1980}. This is because obtaining the original source code from a binary file is an undecidable problem~\cite{Horspool1980}. The cause is that the compiler discards high-level information in the translation process, such as type information, that cannot be recovered in the inverse process.

In the implementation of current decompilers, experts analyze source code snippets and the associated binaries generated by the compiler to identify decompilation patterns. Such patterns associate sequences of assembly instructions with high-level code constructs. These patterns are later included in the implementation of decompilers~\cite{Cifuentes1994,VanEmmerik2007}. The identification of these code generation patterns is not an easy task, because of many factors such as the optimizations implemented by compilers, the high expressiveness degree of high-level languages, the compiler used, the target CPU, and the compilation parameters.

The use of large volumes of source code has been used to create tools aimed at improving software development~\cite{Ortin2016}. This approach has been termed ``big code'' since it applies big data techniques to source code. In the big code area, existing source-code corpora have already been used to create different systems such as JavaScript deobfuscators~\cite{Raychev2015}, automatic translators of C\# code into Java~\cite{Karaivanov2014}, and tools for detecting program vulnerabilities~\cite{Yamaguchi2014}. Probabilistic models are built with machine learning and natural language processing techniques to exploit the abundance of patterns in source code~\cite{Ortin2020}.

Our idea is to use large portions of high-level source code and their related binaries to train machine learning models. Then, these models will help us find code generation patterns not used by current decompilers. The patterns found can be used to improve existing decompilers. Machine learning has already been used for decompilation. Different recurrent neural networks have been used to recover the number and some built-in types of function parameters~\cite{Chua2017}. Extremely randomized trees and conditional random fields have provided good results inferring basic type information~\cite{He2018}. Decompilation has also been tackled with encoder-decoder translation neural networks~\cite{KatzDeborah2018} and with a genetic programming approach~\cite{Schulte2018} (these works are detailed in Section~\ref{section:related-work}).

The main contribution of this paper is the usage of supervised machine learning to improve type information inferred by decompilers. Particularly, we improve the performance of existing decompilers in predicting the types returned by functions in high-level programs. For that purpose, we instrument C source code to label binary patterns with high-level type information. That labeled information is then used to build predictive models. Moreover, the dataset created is used to document the binary patterns found by the classifiers and facilitate its inclusion in the implementation of any decompiler. Our current work is just focused on the Microsoft C compiler for 32-bit Windows binaries, with the default compiler parameters. However, the proposed method could be applied to other languages and compiler settings.

The rest of the paper is structured as follows. Section~\ref{section:motivating-example} describes a motivating example, and related work is discussed in Section~\ref{section:related-work}. Section~\ref{section:system-architecture} describes the architecture of the dataset-generation system. In Section~\ref{section:methodology}, we detail the methodology used, and the evaluation results are presented in Section~\ref{section:evaluation}. Section~\ref{section:extracted-patterns} discusses some interesting patterns discovered with our dataset, and Section~\ref{section:conclusions} presents the conclusions and future work.

\section{A motivating example}
\label{section:motivating-example}

\begin{figure}
\centering
\includegraphics*[angle=0,width=0.8\textwidth]{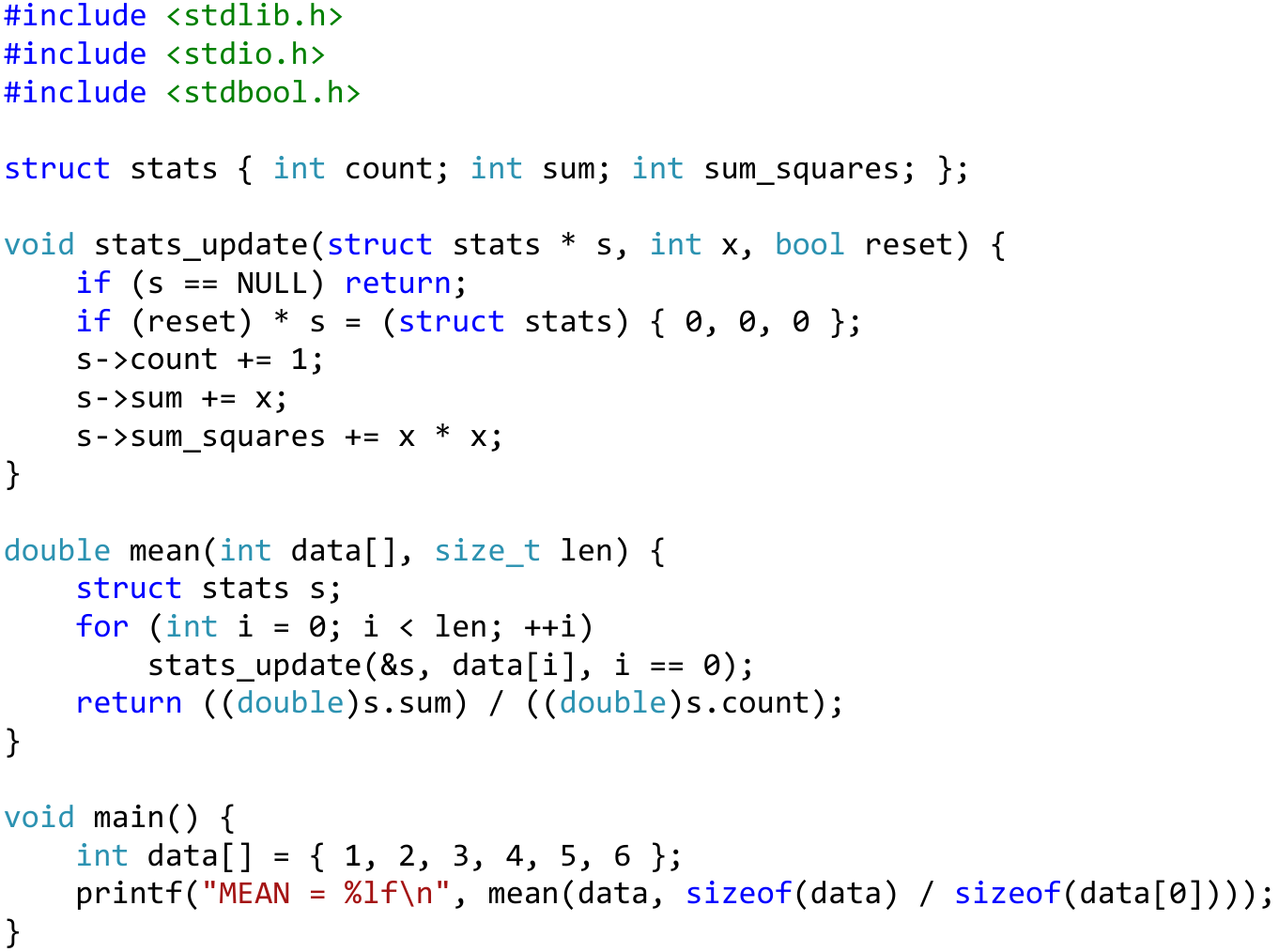}
\caption{Example C source code.}
\label{figure:c-program}
\end{figure}

Figure~\ref{figure:c-program} shows an example C program that we compile with Microsoft \texttt{cl} 32-bit compiler. The generated binary file is then decompiled with four different decompilers, obtaining the following function signatures:

\begin{itemize}
    \item Signature of the \texttt{stats\_update} function:
    \begin{itemize}
        \item IDA Decompiler: \footnotesize\texttt{\textcolor{blue}{int \_\_cdecl} sub\_401000(\textcolor{blue}{int} a1, \textcolor{blue}{int} a2, \textcolor{blue}{char} a3)} 
        \item RetDec: \footnotesize\texttt{\textcolor{blue}{int32\_t} function\_401000(\textcolor{blue}{int32\_t}* a1, \textcolor{blue}{int32\_t} a2, \textcolor{blue}{int32\_t} a3)}
        \item Snowman: \footnotesize\texttt{\textcolor{blue}{void}** fun\_401000(\textcolor{blue}{void}** ecx, \textcolor{blue}{void}** a2, \textcolor{blue}{void}** a3,} \newline \texttt{\textcolor{blue}{void}** a4, \textcolor{blue}{void}** a5, \textcolor{blue}{void}** a6)}
        \item Hopper: \texttt{\textcolor{blue}{int} sub\_401000(\textcolor{blue}{int} arg0, \textcolor{blue}{int} arg1, \textcolor{blue}{int} arg2)}
    \end{itemize}
    \item Signature of the \texttt{mean} function:
    \begin{itemize}
        \item IDA Decompiler: \footnotesize\texttt{\textcolor{blue}{int \_\_cdecl} sub\_4010A0(\textcolor{blue}{int} a1, \textcolor{blue}{unsigned int} a2)}
        \item RetDec: \footnotesize\texttt{\textcolor{blue}{int32\_t} function\_4010a0(\textcolor{blue}{int32\_t} a1, \textcolor{blue}{uint32\_t} a2)}
        \item Snowman: \footnotesize\texttt{\textcolor{blue}{void}** fun\_4010a0(\textcolor{blue}{void}** ecx, \textcolor{blue}{void}** a2, \textcolor{blue}{void}** a3)}
        \item Hopper: \footnotesize\texttt{\textcolor{blue}{int} sub\_4010a0(\textcolor{blue}{int} arg0, \textcolor{blue}{int} arg1)}
    \end{itemize}
\end{itemize}
\bigskip

For both functions, no decompiler infers the correct return type. Although IDA Decompiler, RetDec and Hopper detect the correct number of arguments, most of the parameter types are not correctly recovered by any decompiler\footnote{\texttt{size\_t} and \texttt{uint32\_t} are aliases of \texttt{unsigned int} in a 32-bit architecture.}.

As mentioned, our work is focused on inferring the high-level type returned by functions by analyzing its binary code, which is not an easy task. The reason is that the value is returned to the caller in a registry (\texttt{bool}, \texttt{char}, \texttt{short}, \texttt{int}, \texttt{long}, \texttt{pointer}, and \texttt{struct}\footnote{A \texttt{struct} is commonly returned as a pointer to \texttt{struct} (\textit{i.e.}, its memory address is returned instead of its value).} values are returned in the accumulator; \texttt{long long} in \texttt{edx:eax}; and \texttt{float}, \texttt{double} and \texttt{long double} in the FPU register stack), but the value stored in that registry could be the result of a temporary computation in a function returning \texttt{void}. Therefore, we search for binary patterns before returning from a function and after their invocations, to see if we can recover the high-level return type written by the programmer. 

The problem to be solved is a multi-label classification problem, where the target variable is an enumeration of all the high-level built-in types of the C programming language (including \texttt{void}), plus the type constructors that can be returned\footnote{In C, a function returning an array actually returns a pointer. For Microsoft \texttt{cl}, the \texttt{union} type is actually represented as \texttt{int}, \texttt{long} or \texttt{struct}, depending on its size (explained in Section~\ref{subsection:grouping-types}).} (\texttt{pointer} and \texttt{struct}). In fact, the models we build (Section~\ref{subsubsection:results-high-level-types}) are able to infer the high-level C types returned by the \texttt{stats\_update} and \texttt{mean} functions in Figure~\ref{figure:c-program}.

\section{Related Work}
\label{section:related-work}

\subsection{Type inference}
\label{subsection:type-inference}

There are some research works aimed at inferring high-level type information from binary code. Chua \textit{et al.}~\cite{Chua2017} use recurrent neural networks (RNN)~\cite{Rumelhart1986} to detect the number and type of function parameters. First, they transform each instruction into word embeddings (256 double values per instruction)~\cite{Bengio2003}. Then, a sequence of instructions (vectors) is used to build 4 different RNNs for counting caller arguments and function parameters, recovering types of caller arguments, and recovering types of function parameters. In this work, they infer seven different types: \texttt{int}, \texttt{float}, \texttt{char}, \texttt{pointer}, \texttt{enum}, \texttt{union} and \texttt{struct}. They only consider \texttt{int} for integer values and \texttt{float} for real ones. They achieved 84\% accuracy for parameter counting and 81\% for type recovery.

He \textit{et al.}~\cite{He2018} build a prediction system that takes as an input a stripped binary and outputs a new binary with debug information that includes type information. They combine extremely randomized trees (ERT) with conditional random fields (CRFs)~\cite{Lafferty2001}. The ERT model aims to extract identifiers. Although identifiers are always mapped to registers and memory offsets, not every register and memory offset stores identifiers. Then, the CRF model predicts the name and type of the identifiers discovered by ERT. They use a maximum a posteriori (MAP) estimation algorithm to find a globally optimal assignment. This tool handles 17 different types, but it lacks floating-point types support. This is because the library used to handle the assembly code, Binary Analysis Platform (BAP)~\cite{Brumley2011}, does not support floating-point instructions. Their system achieves 68.8\% precision and 68.3\% recall.

Some works are not focused on type inference exclusively, but they undertake decompilation as a whole, including type inference. The works in~\cite{KatzDeborah2018, KatzOmer2019, Fu2019} propose different systems based on neural machine translation~\cite{Kalchbrenner2013, Sutskever2014}. They use RNNs with an encoder-decoder scheme to learn high-level code fragments from binary code, for a given compiler.

Katz (Deborah) \textit{et al.}~\cite{KatzDeborah2018} use RNN models for snippet decompilation with additional post-processing techniques. They tokenize the binary input with a byte-by-byte approach and the output with a C lexer. C tokens are ranked by their frequency and replaced by the ranking position. Less frequent tokens (below a frequency threshold) are replaced by a common number to minimize the vocabulary size. This transformation reduces the number of tokens, speeding up the training of the RNN. For the binary input, a language model is created, and byte embeddings are found for the binary information. Once the encoder-decoder scheme is trained, translation from binary code into C tokens is performed. The final step is to apply several post-processing transformations, such as deleting extra semicolons, adding missing commas, and balancing brackets, parenthesis, and curly braces.

The previous work was later modified by Katz (Omer) \textit{et al.} to reduce the compiler errors in the output C code~\cite{KatzOmer2019}. In this previous work, most of the output C code could not be compiled because errors were found Therefore, they modified the decoder so that it produces prefixed templates to be filled. This idea is inspired by delexicalization~\cite{Henderson2014}. Delexicalized features are n-gram features where references to a particular slot or value are replaced with a generic symbol. In this way, locations in the output C source code are substituted with placeholders. After the translation takes place, those placeholders are replaced with values and constants taken from the binary input, improving the code recovery process up to 88\%.

Coda~\cite{Fu2019} is an end-to-end neural-based framework for code decompilation. First, Coda employs an instruction type-aware encoder and a tree decoder for generating an abstract syntax tree (AST), using attention feeding during the code sketch generation stage. Afterwards, it updates the code sketch using an iterative error-correction machine, guided by an ensembled neural error predictor. An approximate candidate is found first. Then, the candidate is fixed to produce a compilable program. Evaluation results show that Coda achieves 82\% program recovery accuracy on unseen binary samples.

Schulte \textit{et al.}~\cite{Schulte2018} propose genetic programming to generate readable C source code from compiled binaries. Taking binary code as the input, an evolutionary search seeks a combination of source code excerpts from a big code database. That source code is compiled into an executable, which should be byte-equivalent to the original binary. The decompiled source code reproduces the behavior, both intended and unintended, of the original binary. As they use evolutionary search, decompilation time can vary dramatically between executions.

Mycroft~\cite{Mycroft1999} proposes a type reconstruction algorithm that uses unification to recover types from binary code. Mycroft starts by transforming the binary code into a register transfer language (RTL) intermediate representation. The RTL representation is then transformed into a single static assignment (SSA) form to undo some optimizations performed by the compiler. Then, each code instruction is used to generate constraints about the type of its operands, regarding their use. Those constraints are used to recover types by applying a modified version of Milner's \textit{W} algorithm~\cite{Milner1978}. In this variation, any constraint violation causes type reconstruction (recursive structs and unions) instead of premature termination. This work does not discuss stack-based variables, only register-based ones.

All these works represent complement methods, rather than alternatives, to the established approaches used in decompiler implementations, such as value-based, flow-based and memory-based analyses~\cite{Caballero2016}.

\subsection{Other uses of machine learning for code reversing}
\label{subsection:other-uses-ml}

Apart from recovering high-level type information from binaries, other works use machine learning for different code reversing purposes~\cite{Xue2019}. Rosenblum \textit{et al.}~\cite{Rosenblum2008} use CRFs to detect function entry points (FEPs). They use n-grams of the generalized instructions surrounding FEPs, together with a call graph representing the interaction between FEPs. The FEP detection problem consists in finding the boundaries of each function in the binary code. CRFs allow using both sources of information together. Since standard inference methods for CRFs are expensive, they speed up training with approximate inference and feature selection. Nonetheless, feature selection took 150 days of computation on 1171 binaries. This approach does not seem to be tractable for big code scenarios.

Bao \textit{et al.}~\cite{Bao2014} utilize weighted prefix trees (or weighted tries)~\cite{Briandais1959} to detect FEPs, considering generalized instructions as tree nodes. Once trained, each node represents the likelihood that the sequence from the root node to the current node will be an FEP. They trained the model with 2064 binaries in 587 computing hours, obtaining better results than~\cite{Rosenblum2008}. The approach of Shin \textit{et al.}~\cite{Shin2015} uses RNNs for the same problem. The internal feedback loops of RNNs makes them suitable to handle sequences of bytes. This approach reduces training time to 80 computing hours using the same dataset as Bao \textit{et al.}~\cite{Bao2014}, while performing slightly better.

Rosenblum \textit{et al.}~\cite{Rosenblum2010} use CRFs to detect the compiler used to generate binary files. Binaries frequently exhibit gaps between functions. These gaps may contain data such as jump tables, string constants, regular padding instructions and arbitrary bytes. While the content of those gaps sometimes depends on the functionality of the program, they sometimes denote compiler characteristics. A CRF model is built to exploit these differences among binaries and infer the compiler used to generate the binaries.  They later extend this idea to detect different compiler versions and optimization levels~\cite{Rosenblum2011}.

Malware detection is another field related to binary analysis where machine learning has been used~\cite{Ucci2017}. Alazab \textit{et al.}~\cite{Alazab2010} separate malware from benign software by analyzing the sequence of Windows API calls performed by the program. First, they process the binaries to extract all the Windows API invocations. Then, those sequences are vectorized and used to build eight different supervised machine learning models. Those models are finally evaluated, finding that support vector machine (SVM) with a normalized poly-kernel classifier is the method with the best results. SVM achieves 98.5\% accuracy.

Rathore \textit{et al.}~\cite{Rathore2018} detect malware by analyzing opcodes frequency. They use various machine learning algorithms and deep learning models. In their experiments, random forest outperforms deep neural networks. Static analysis of the assembly code is used to generate multi-dimensional datasets representing opcode frequencies. Different feature selection strategies are applied to reduce dimensionality. They collect binaries from different sources, selecting 11,688 files with malware and 2,819 benign executables. The dataset is balanced with adaptive synthetic (ADASYN) sampling. Random forest obtained 99.78\% accuracy. 

\section{System architecture}
\label{section:system-architecture}

\begin{figure}[ht]
\centering
\includegraphics*[angle=0,width=\textwidth]{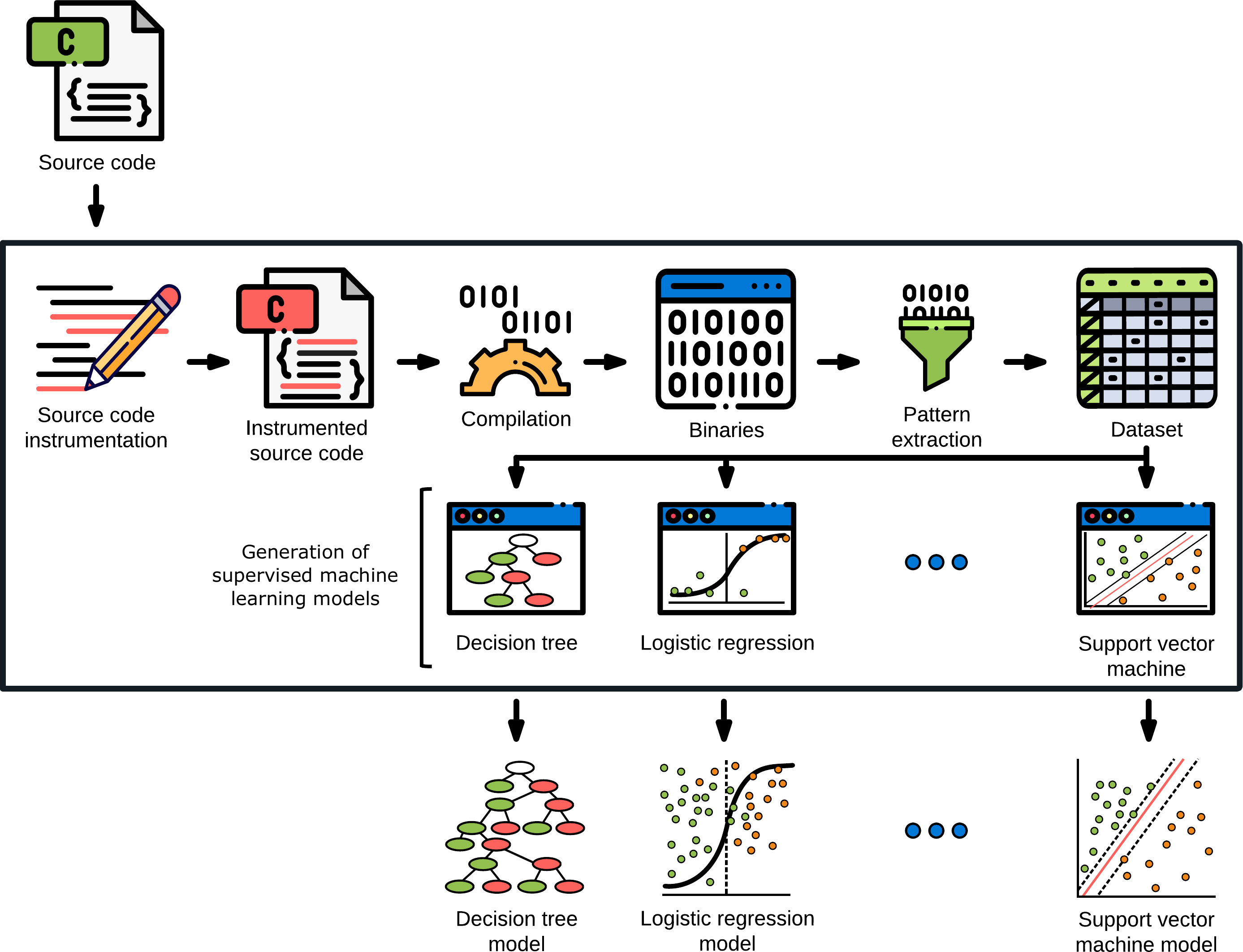}
\caption{System architecture.}
\label{figure:system-architecture}
\end{figure}

Figure~\ref{figure:system-architecture} shows the architecture of our system. It receives C source code as an input and generates different machine learning models. Each model is aimed at predicting the high-level type returned by the functions in a program, by just receiving its binary representation.

The system starts instrumenting C source code. This step embeds annotations in the input program to allow associating high-level constructs to their binary representation (Section~\ref{subsection:instrumentation})~\cite{Escalada2017}. After that, the instrumented source code is compiled to obtain the binaries. A pattern extraction process analyzes the binaries looking for the annotations, collecting the set of binary patterns related to each function invocation and \texttt{return} expression. Finally, the resulting dataset is created, where binary patterns are associated with the return type of each function.

Table~\ref{table:dataset} shows the simplified structure of datasets generated by our system. Each row (individual, instance or sample) represents a function from the input C source code. Each column (feature or independent variable) but the last one represents a binary pattern found by the pattern extraction process. For example, the first pattern in Table~\ref{table:dataset} is the assembly code for a \texttt{return} expression of some functions returning an \texttt{int} literal. That value is moved to the \texttt{eax} 32-bit register, followed by the code that all functions use to return to the caller (\textit{callee\_epilogue}). The second feature is the binary code of a function invocation (\textit{caller\_epilogue}) followed by a \texttt{cwde} instruction. Since \texttt{cwde} converts the signed integer representation from \texttt{ax} (16 bits) to \texttt{eax} (32 bits), the target class in the dataset is set to the \texttt{short} high-level type (Section~\ref{subsection:pattern-extraction} explains how the dataset is built).

After creating the dataset, the system trains different classifiers following the methodology described in Section~\ref{section:methodology}. The forthcoming subsections detail each of the modules in the architecture.

\begin{table}
\centering
\footnotesize
\setlength{\tabcolsep}{6pt} 
\begin{tabular}{|c|c|c|c|c|}
\hline
& \multicolumn{1}{l|}{\textbf{(RET)}} & \multicolumn{1}{l|}{\textbf{(POST CALL)}} & & \\
& \multicolumn{1}{l|}{\texttt{mov eax}, \textit{literal}} & \multicolumn{1}{l|}{\textit{caller\_epilogue}} & ... & Return type \\
& \multicolumn{1}{l|}{\textit{callee\_epilogue}} & \multicolumn{1}{l|}{\texttt{cwde}} & & \\
\hline
func\textsubscript{1} & \cellcolor[gray]{.8}\hfil1 & \cellcolor[gray]{.8}\hfil0 & \cellcolor[gray]{.8}... & \texttt{int} \\
\hline
func\textsubscript{2} & \cellcolor[gray]{.8}\hfil0 & \cellcolor[gray]{.8}\hfil1 & \cellcolor[gray]{.8}... & \texttt{short} \\
\hline
... & \cellcolor[gray]{.8}\hfil... & \cellcolor[gray]{.8}\hfil... & \cellcolor[gray]{.8}... & ... \\
\hline
func\textsubscript{n-1} & \cellcolor[gray]{.8}\hfil0 & \cellcolor[gray]{.8}\hfil1 & \cellcolor[gray]{.8}... & \texttt{short} \\
\hline
func\textsubscript{n} & \cellcolor[gray]{.8}\hfil0 & \cellcolor[gray]{.8}\hfil0 & \cellcolor[gray]{.8}... & \texttt{double} \\
\hline
\end{tabular}
\caption{Example dataset generated by our system.}
\label{table:dataset}
\end{table}

\subsection{Instrumentation}
\label{subsection:instrumentation}

As mentioned, much high-level information is discarded in the compilation process. One example is the association between a high-level \texttt{return} statement and its related assembler instructions. There is not a direct way to identify the binary code generated for a \texttt{return} statement. For this reason, our instrumentation process includes no-operational code around some syntactic constructs in the input C program. The instrumented code does not change the semantics of the program, but help us find the binary code generated for different high-level code snippets.

\begin{figure}
\centering
\includegraphics*[angle=0,width=0.9\textwidth]{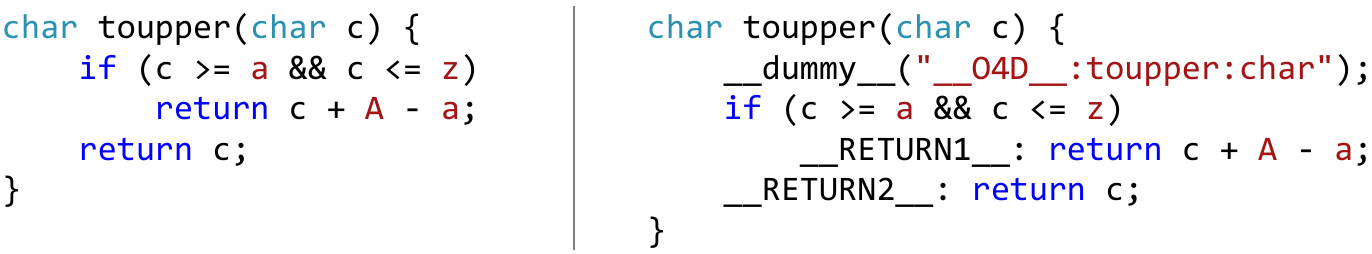}
\caption{Original C source code (left) and its instrumented version (right).}
\label{figure:struct-code}
\end{figure}

The left-hand side of Figure~\ref{figure:struct-code} shows an example of the original C function, and its right-hand side presents the instrumented version. The function \texttt{\_\_dummy\_\_} performs no action. Its invocation is added to provide information about the name and return type of the high-level function in the binary code.

The example in Figure~\ref{figure:struct-code} also shows the instrumentation included to delimit the binary code of the expressions to be returned by a function. To this end, different \texttt{\_\_RETURN\textit{n}\_\_} labels are added before every \texttt{return} statement. The end of the returned expression is delimited by the \texttt{retn} assembly instruction. 

\subsection{Compilation}
\label{subsection:compilation}

The instrumented C source code is compiled to obtain the binaries. In this work, we only use the Microsoft \texttt{cl} compiler to build native applications for Intel x86 32-bit microprocessors. The compilation parameters are the default ones. We plan to apply the method proposed in this article with other compilers, parameters and microprocessors.

\subsection{Pattern extraction}
\label{subsection:pattern-extraction}

The pattern extraction module performs three processes: extraction of binary chunks, pattern generalization and dataset generation.

\subsubsection{Extraction of binary chunks}
\label{subsubsection:extraction-binary-chunks}

\begin{figure}
  \centering
  \begin{minipage}{.3\linewidth}
    \begin{algorithm}[H]
        \footnotesize
        \SetAlgoVlined
        \SetKwBlock{Chunk}{binary chunk}{end}
        \Chunk{
        \hspace{8ex}\\
        \texttt{\_\_RETURN\textit{n}\_\_:}\\
        \hspace{8ex}\textit{instr\textsubscript{n}}\\
        \hspace{8ex}\textit{instr\textsubscript{n-1}}\\
        \hspace{8ex}\textit{...} \\
        \hspace{8ex}\textit{instr\textsubscript{2}}\\
        \hspace{8ex}\textit{instr\textsubscript{1}}\\
        \hspace{8ex}\texttt{retn}\\
        }
    \end{algorithm}
  \end{minipage}
  \caption{Different RET patterns taken from a binary chunk.}
  \label{figure:ret-pattern}
\end{figure}

The first task is to extract the binary chunks associated with every \texttt{return} and function invocation statements. For \texttt{return} statements, we get the binary code between a \texttt{\_\_RETURN\textit{n}\_\_} label and the next \texttt{retn} instruction. Since we do not know how many instructions are sufficient to predict the return type (one high-level expression may produce many low-level instructions), we generate different binary patterns with a growing number of binary instructions before \texttt{retn}. Figure~\ref{figure:ret-pattern} shows an example binary chunk for a given \texttt{return} statement. The smallest pattern goes from \textit{instr\textsubscript{1}} to \texttt{retn}. The following contains \textit{instr\textsubscript{2}}, \textit{instr\textsubscript{1}} and \texttt{retn}. The biggest one contains the whole chunk, from \textit{instr\textsubscript{n}} to \texttt{retn}. We call these types of binary sequences RET patterns.

The other kind of binary chunks we retrieve are the sequences of instructions after each function invocation. We consider the \texttt{call} instruction, the optional \texttt{add esp}, \textit{literal} instruction used to pop the invocation arguments from the stack, and the following assembly instructions. We call these binary sequences POST CALL patterns.

As for RET patterns, we first created different POST CALL patterns with an increasing number of instructions after function \texttt{call} and stack restoration. However, after evaluating the classification models built from the datasets, we realized that only the first assembly instruction was used by the models to predict the returned type. Therefore, we only consider the first instruction after stack restoration in POST CALL patterns.

\subsubsection{Pattern generalization}
\label{subsubsection:pattern-generalization}

If we use the exact representation of binary instructions as features in the dataset, the machine learning algorithms will consider instructions such as \texttt{mov eax, 32} and \texttt{mov eax, 33} to be different. If we do not generalize such instructions to represent the same feature, the predictive models will not be accurate enough and the information extracted from them (Section~\ref{section:extracted-patterns}) will not be understandable. For instance, the previous assembly instructions are generalized by our system to the two following patterns: \texttt{mov eax}, \textit{literal} and \texttt{mov} \textit{reg}, \textit{literal} (one instruction may be represented by different generalizations).

Table~\ref{table:basic-generalization-patterns} shows different generalization examples implemented by our system. \textit{Operand} is the most basic generalization, which groups some types of operands, such as literals, addresses and indirections. \textit{Mnemonic} generalizations group instructions with similar functionalities, such as \texttt{mov}, \texttt{movzx} y \texttt{movsx}. The last type of generalizations, \textit{Sequence}, clusters sequences of instructions that appear multiple times in the binary code. For instance, \textit{callee\_epilogue} and \textit{caller\_epilogue} appear, respectively, before returning one expression and after invoking a function.

\begin{table}[H]
\centering
\footnotesize
\begin{tabular}{cll}
\toprule
& Instruction sequences & Generalized pattern \\
\midrule
\multirow{9}{*}{\begin{sideways}{Operand}\end{sideways}}
& \texttt{sub al, 1} & \texttt{sub al,} \textit{literal} \\
& \texttt{mov ecx, [ebp+var\_1AC8]} & \texttt{mov ecx, [ebp+}\textit{literal}\texttt{]} \\
& \texttt{mov ecx, [ebp+var\_1AC8]} & \texttt{mov ecx, [}\textit{reg}\texttt{]} \\
& \texttt{push offset \$SG25215} & \texttt{push} \textit{address} \\
& \texttt{movsd xmm0, ds:\_\_real@43e2eb565391bf9e} & \texttt{movsd xmm0,} \textit{*address} \\
& \texttt{jmp loc\_22F} & \texttt{jmp} \textit{offset} \\
& \texttt{mov cx, [ebp+eax*2+var\_10]} & \texttt{mov cx, [ebp+eax*}\textit{literal\textsubscript{1}}\texttt{+}\textit{literal\textsubscript{2}}\texttt{]} \\
& \texttt{mov cx, [ebp+eax*2+var\_10]} & \texttt{mov cx, [ebp+}\textit{reg}\texttt{*}\textit{literal\textsubscript{1}}\texttt{+}\textit{literal\textsubscript{2}}\texttt{]} \\
& \texttt{mov cx, [ebp+eax*2+var\_10]} & \texttt{mov ecx, [}\textit{reg}\texttt{]} \\
\midrule
\multirow{3}{*}{\begin{sideways}{Mnem.}\end{sideways}}
& \texttt{movzx ecx, [ebp+var\_A]} & \textit{mov} \texttt{ecx, [ebp+var\_A]} \\
& \texttt{movsx ecx, \_global\_var\_1234} & \textit{mov} \texttt{ecx, \_global\_var\_1234} \\
& \texttt{mov [eax], edx} & \textit{mov} \texttt{[eax], edx} \\
\midrule
\multirow{20}{*}{\begin{sideways}{Sequence}\end{sideways}}
& \texttt{pop esi} & \multirow{5}{*}{\textit{callee\_epilogue}} \\ [-0.25em]
& \texttt{pop edi} & \\ [-0.25em]
& \texttt{mov esp, ebp} & \\ [-0.25em]
& \texttt{pop ebp} & \\ [-0.25em]
& \texttt{retn} & \\
\addlinespace
& \texttt{mov esp, ebp} & \multirow{3}{*}{\textit{callee\_epilogue}} \\ [-0.25em]
& \texttt{pop ebp} & \\ [-0.25em]
& \texttt{retn} & \\
\addlinespace
& \texttt{pop ebp} & \multirow{2}{*}{\textit{callee\_epilogue}} \\ [-0.25em]
& \texttt{retn} & \\
\addlinespace
& \texttt{call \_func56} & \multirow{2}{*}{\textit{caller\_epilogue}} \\ [-0.25em]
& \texttt{add esp, 4} & \\
\addlinespace
& \texttt{call \_proc2} & \textit{caller\_epilogue} \\
\addlinespace
& \texttt{mov eax, 0 } & \multirow{3}{*}{\textit{mov\_chain}(\texttt{[ebp+var\_8], ebx, eax, 0})} \\ [-0.25em]
& \texttt{mov ebx, eax} & \\ [-0.25em]
& \texttt{mov [ebp+var\_8], ebx} & \\ [-0.25em]
\addlinespace
& \texttt{ja loc\_D9B6B} & \multirow{4}{*}{\textit{bool\_cast}(\texttt{[ebp+var\_10]})} \\ [-0.25em]
& \texttt{mov [ebp+var\_10], 1} & \\ [-0.25em]
& \texttt{jmp loc\_D9B72} & \\ [-0.25em]
& \texttt{mov [ebp+var\_10], 0} & \\
\bottomrule
\end{tabular}
\caption{Generalization examples made by our system. \textit{reg} variables represent registers, \textit{literal} integer literals, \textit{address} absolute addresses, \textit{*address} absolute addresses dereferencing and \textit{offset} relative addresses.}
\label{table:basic-generalization-patterns}
\end{table}

\subsubsection{Dataset creation}
\label{subsubsection:dataset-creation}

After pattern generalization, the datasets are created before training the models (Table~\ref{table:dataset}). Each cell in the dataset indicates the occurrence of each pattern (column) in every single function (row) in the program. RET patterns are associated with the function bodies, but POST CALL patterns are related to the invoked function. For example, if a function \texttt{f} is invoked in the body of a function \texttt{g}, the POST CALL pattern will be associated with the row representing the function \texttt{f}, not \texttt{g}.

Finally, the return type (target) of each function should be added to the dataset. Our system gets that information from the string parameter passed to the \texttt{\_\_dummy\_\_} function added in the instrumentation process.

\section{Methodology}
\label{section:methodology}

\subsection{Dataset}
\label{subsection:dataset}

\begin{table}
\centering
\footnotesize
\begin{threeparttable}
\begin{tabular}{lrrl}
\toprule
Project & Functions & LoC & Description \\
\midrule
arcadia & 121 & 3,590 & Implementation of Arc, a Lisp dialect\tnote{a}. \\
bgrep & 5 & 252 & Grep for binary code\tnote{b}. \\
c\_ray\_tracer & 52 & 1,063 & Simple ray tracer\tnote{c}. \\
\multirow{2}{*}{jansson} & \multirow{2}{*}{176} & \multirow{2}{*}{7,020} & Library for encoding, decoding and manipulating JSON 
      \\ & & & data \tnote{d}. \\
\multirow{2}{*}{libsodium} & \multirow{2}{*}{642} & \multirow{2}{*}{35,645} & Library for encryption, decryption, signatures and password \\
 & & & hashing\tnote{e} .\\
lua 5.2.3 & 820 & 14,588 & The Lua programming language\tnote{f}. \\
masscan & 496 & 26,316 & IP port scanner\tnote{g}. \\
slre & 17 & 564 & Regular expression library\tnote{h}. \\
\midrule
Total & 2,329 & 89,038 & \\
\bottomrule
\end{tabular}
\begin{tablenotes}
\scriptsize
\item[a]\url{https://github.com/kimtg/arcadia}
\item[b]\url{https://github.com/elektrischermoench/bgrep}
\item[c]\url{https://web.archive.org/web/20150110171135/http://patrickomatic.com/c-ray-tracer}
\item[d]\url{https://github.com/akheron/jansson}
\item[e]\url{https://github.com/jedisct1/libsodium}
\item[f]\url{https://lua.org/download.html}
\item[g]\url{https://github.com/robertdavidgraham/masscan}
\item[h]\url{https://github.com/cesanta/slre}
\end{tablenotes}
\end{threeparttable}
\caption{Open-source C projects used.}
\label{table:projects-used}
\end{table}

Table~\ref{table:projects-used} shows the different open source C projects we used to create the dataset. Although they sum 2329 functions and 89,038 lines of code, they do not represent sufficient data to infer the types returned by functions. Unfortunately, there are not many open-source C (not C++) programs compilable with Microsoft \texttt{cl} compiler. Moreover, we want to generate a dataset with a balanced number of instances for each return type, so the number of functions in Table~\ref{table:projects-used} would even be lower. To increase the size of our dataset, we developed an automatic C source code generator called Cnerator~\cite{cnerator}.

Cnerator generates valid C programs to be compiled with any standard ANSI C compiler. One of its modes allows the user to specify the number of functions to be created. It also allows specifying different probabilities of the synthetic code to be generated, such as the average number of statements in a function, expression types, number and types of local variables, and the kind of syntactic constructs to be generated, among others. The generated programs fulfill the type rules of the C programming language, so they are compiled without errors. Using those probabilities, we make Cnerator generate synthetic programs with unusual language constructs, which programmers rarely use. This facilitates the creation of datasets covering a wide range of C programs.

The final dataset comprises the source code of the ``real'' projects in Table~\ref{table:projects-used} plus the synthetic code generated by Cnerator. On one hand, the synthetic code provides a huge number of functions, a balanced dataset, and all the language constructs we want to include. On the other hand, real projects increase the probability of those patterns that real programmers often use (\textit{e.g.}, most C programmers use \texttt{int} expressions instead of \texttt{bool} for Boolean operations). This combination of real and synthetic source code improves the predictive capability of our dataset.

\subsection{Dataset size}
\label{subsection:dataset-size}

Since Cnerator~\cite{cnerator} allows us to generate any number of functions (individuals), we should find out the necessary number of synthetic functions to include in our dataset in order to build models with the highest performance. To determine this number, we conduct the following experiment. We start with a balanced dataset with 100 functions for each return type. Then, we build different classifiers (see Section~\ref{subsection:classification-algorithms}) and evaluate their accuracy. Next, we add 1000 more synthesized functions to the dataset, re-build the classifiers and re-evaluate them. This process is repeated until the accuracy of classifiers converge. To detect this convergence, we compute the coefficient of variation (CoV) of the last accuracies, stopping when that coefficient is lower than~2\%.

\subsection{Classification algorithms}
\label{subsection:classification-algorithms}

We use the following 14 classifiers from scikit-learn~\cite{scikit-learn}: logistic regression (\texttt{Lo\-gis\-tic\-Regression}), perceptron (\texttt{Perceptron}), multilayer perceptron (\texttt{MLP\-Classifier}), Ber\-nou\-lli na\"ive Bayes (\texttt{Bernoulli\-NB}), Gaussian na\"ive Bayes (\texttt{Gaussian\-NB}), multinomial na\"ive Bayes (\texttt{Multinomial\-NB}), decision tree (\texttt{Decision\-Tree\-Classifier}), random forest (\texttt{Random\-Forest\-Classifier}), extremely randomized trees (\texttt{Extra\-Trees\-Classifier}), support vector machine (\texttt{SVC}), linear support vector machine (\texttt{Linear\-SVC}), AdaBoost (\texttt{AdaBoost\-Classifier}), gradient boosting (\texttt{Gradient\-Boosting\-Cla\-ssi\-fier}), k-nearest neighbors (\texttt{KNeighbors\-Classifier}).

In the process described in Section~\ref{subsection:dataset-size} to find the optimal size of the dataset, we use a stratified and randomized division of the dataset (\texttt{Stratified\-Shuffle\-Split} class in scikit-learn). 80\% of the instances in the dataset are used for training and the remaining 20\% for testing. Since each classifier has a different optimal size, we choose the greatest optimal size among all the classifiers (results are shown in Sections~\ref{subsubsection:data-size} and~\ref{subsection:high-level-types}).

\subsection{Feature selection}
\label{subsection:feature-selection}

As mentioned, our system generates a lot of features because, for each pattern, different generalizations are produced. Therefore, a feature selection mechanism would be beneficial to avoid the curse of dimensionality and enhance the generalization property of the classifiers. Consequently, after creating the datasets with the optimal size, we select the appropriate features to build each model.

We follow a wrapper approach~\cite{Kohavi1997} for each classification algorithm, evaluating different feature selection techniques and selecting the one that obtains the best classification performance. If performance differences between two feature selection techniques are not significantly different, we choose the one that selects a lower number of features (see the results in Sections~\ref{subsubsection:feature-selection} and~\ref{subsection:high-level-types}). The performance of feature selection is evaluated with the training dataset (80\% of the original one) using 3-fold stratified cross-validation (\texttt{Stratified\-Shuffle\-Split}).

We use both recursive feature elimination (\texttt{RFECV}) and selection from a model (\texttt{Select\-From\-Model}) feature selection techniques. Given an external estimator that assigns weights to features, \texttt{RFECV} selects the features by recursively considering smaller feature sets. \texttt{Select\-FromModel} discards the features that have been rejected by some other classifiers like tree-based ones. Four \texttt{Select\-FromModel} configurations were used: random forest and extremely randomized trees as classifiers to select the features; and the mean and median thresholds to filter features by their importance score.

\subsection{Hyperparameter tuning}
\label{subsection:hyperparameter-tuning}

After feature selection, we tune the hyperparameters of each model. To that end, we use \texttt{Grid\-Search\-CV}, which performs an exhaustive search over the specified hyperparameter values. Similar to the feature selection process, the 80\% training set is used to validate the hyperparameters with 3-fold stratified cross-validation (\texttt{Stratified\-Shuffle\-Split}).

The final hyperparameters selected for each classifier are available at~\cite{paperwebpage}. For the multilayer perceptron neural network, we use a single hidden layer with 100 units, the sigmoid activation function, Adam optimizer, and softmax as the output function.

\subsection{Evaluation of model performance}
\label{subsection:evaluation}

After feature selection and hyperparameter tuning, we create and evaluate different models (one for each algorithm in Section~\ref{subsection:classification-algorithms}) to predict the type returned by functions. As mentioned, the dataset has real functions coded by programmers, and synthetic ones generated by Cnerator. We consider these two types of code to define three different methods to evaluate the performance of the classifiers:

\begin{enumerate}

	\item \label{item:first}Mixing real and synthetic functions. This is the simplest evaluation method, where real and synthetic functions are merged in the dataset. 80\% of them are used for training and the remaining 20\% for testing, so sets have the same percentage of real and synthesized functions. The training and test sets are created with stratified randomized selection, making all the classes to be equally represented in both sets. 

	\item \label{item:second}Estimate the necessary number of real functions for training. Since we have lots of synthetic functions, we want to estimate to what extent synthetic programs can be used to classify code written by real programmers. We first create a model only with all the synthetic functions generated to determine the dataset size (Section~\ref{subsection:dataset-size}) and test it with the real functions. Then, we include 1\% of real functions in the training dataset, rebuild and retest the models, and see the accuracy gain. This process stops when the CoV is lower than 1\% for the last 10 accuracies. The obtained percentage of real functions in the training set indicates how many real functions are necessary to build accurate predictive models (32\% for the experiment in Section~\ref{subsection:grouping-types} and 48\% for that in Section~\ref{subsection:high-level-types}). Decision tree was the classifier used to estimate this value.

	\item \label{item:third}Prediction of complete real programs. This evaluation method measures prediction for source code written by programmers whose code has not been included in the test dataset. One real program is used to build the test dataset, and no functions of that program are used for training. In this case, we evaluate whether our system is able to predict return types for unknown programming styles.

\end{enumerate}

\subsection{Selected decompilers}
\label{subsection:selected-decompilers}

We compare our models with the following existing decompilers:

\begin{itemize}

	\item[--] IDA Decompiler~\cite{hex-rays-decompiler}. This is a plugin of the commercial Hex-Rays IDA disassembler~\cite{hex-rays-ida}. This product is the result of the research works done by Ilfak Guilfanov~\cite{Guilfanov2001, Guilfanov2008}. This tool is the current \textit{de facto} standard in software reverse engineering.

	\item[--] RetDec~\cite{retdec}. An open-source decompiler developed initially by K\v{r}oustek~\cite{Kroustek2015}, currently maintained by the AVAST company. It can be used as a standalone application or as a Hex-Rays IDA plugin. To avoid the influence of the Hex-Rays IDA decompiler, we use the standalone version.

	\item[--] Snowman~\cite{snowman}. Open-source decompiler based on the TyDec~\cite{Troshina2009} and SmartDec~\cite{Fokin2011} proposals. Similar to RetDec, it can also be used as a standalone application or as a Hex-Rays IDA plugin. We use the standalone version.

	\item[--] Hopper~\cite{hopper}. A commercial decompiler developed by Cryptic Apps. Although it is mainly focused on decompiling Objective-C, it also provides C decompilation of any Intel x86 binary.

\end{itemize}

We also considered other alternatives that we finally did not include in our evaluation. DCC~\cite{Cifuentes1994} and DISC~\cite{disc} decompilers do not work with modern executables. The former is aimed at decompiling MS-DOS binaries, while the latter only decompiles binaries generated with TurboC. Boomerang~\cite{VanEmmerik2007} and REC~\cite{rec} are no longer maintained. Phoenix~\cite{Schwartz2013} is built on the top of the Binary Assembly Platform~\cite{Brumley2011}, which lacks support for floating-point instructions. Lastly, we could not find the implementations of the DREAM~\cite{Yakdan2015} and DREAM++~\cite{Yakdan2016} decompilers.

\subsection{Data analysis}
\label{subsection:data-analysis}

For each classifier, we compute its performance following the three different evaluation methods described in Section~\ref{subsection:evaluation}. We repeat the training plus testing process 30 times, computing the mean, standard deviation and 95\% confidence intervals of model accuracies. This allows us to compare accuracies of different models, checking whether two evaluations are significantly different when their two 95\% confidence intervals do not overlap
~\cite{Georges2007}. Figures showing model accuracies (Figures~\ref{figure:size-accuracy-graph} and~\ref{figure:all-accuracy-graph}) display the 95\% confidence intervals as whiskers.

In a balanced multi-class classification, overall precision and recall are usually computed as the average values for each class. These aggregate metrics are called macro-precision and macro-recall~\cite{Yang1999}. Likewise, macro-F\textsubscript{1}-score can be computed as the average of per-class F\textsubscript{1}-score~\cite{Yang1999}, or as the harmonic mean of macro-precision and macro-recall~\cite{Sokolova2009}. We use the first alternative because it is less sensitive to error type distribution~\cite{Opitz2019}. For the sake of brevity, we use precision, recall and F\textsubscript{1}-score to refer to the actual macro-precision, macro-recall and macro-F\textsubscript{1}-score measurements.

We run all the code in a Dell PowerEdge R530 server with two Intel Xeon E5-2620 v4 2.1 GHz microprocessors (32 cores) with 128GB DDR4 2400 MHz RAM, running an updated version of Windows 10 for 64 bits.

\section{Evaluation}
\label{section:evaluation}

In the assembly language, the concept of type is related to the size of values more than to the operations that can be done with those values. For example, the integer \texttt{add} instruction works with 8 (\texttt{ah}), 16 (\texttt{ax}) and 32 bits (\texttt{eax}), but it is not checked whether the accumulator register is actually holding an integer. For this reason, in this paper we evaluate two different kinds of models: those considering types by their size and representation (Section~\ref{subsection:grouping-types}), and those considering types by the operations they support---\textit{i.e.}, high-level C types (Section~\ref{subsection:high-level-types}).

The first kind of models predicts return types when they have different size or representation. In this way, these models separate \texttt{short} (2 bytes) from \texttt{int} (4 bytes). They also tell the difference between \texttt{int} and \texttt{float}, because, even though their size is 4 bytes, their representations are different (integer and real). On the contrary, \texttt{char} and \texttt{bool} are not distinguished in the first kind of models since they both are 1-byte sized and hold integer values (C does not provide different operations for \texttt{char} and \texttt{bool}).

After building and evaluating these type-by-size-and-representation models (Section~\ref{subsection:grouping-types}), we define additional mechanisms to distinguish among types with similar sizes to improve our models. Thus, Section~\ref{subsection:high-level-types} shows additional generalization patterns to improve the classification of high-level return types. With those enhancements, our models improve the differentiation among types with the same size such as \texttt{char} and \texttt{bool}, and \texttt{int} and \texttt{pointer}\footnote{Note that, in assembly, there is no difference in the \emph{representation} of integers, characters, Booleans and pointers, because, for the microprocessor, all of them hold integer values.}.

\subsection{Grouping types by size and representation}
\label{subsection:grouping-types}

In binary code, the value returned by a function is passed to the caller via registers. CPUs have different kinds of registers depending on their sizes and representations (integer or a floating-point number). In the particular case of Intel x86, registers can hold integer values of 8-, 16- or 32-bit, and 32- or 64-bit floating-point numbers.

\begin{table}
\centering
\footnotesize
\begin{tabular}{ll}
\toprule
Target & C high-level type \\
\midrule
\texttt{INT\_1} & \texttt{bool} and \texttt{char} \\
\texttt{INT\_2} & \texttt{short} \\
\texttt{INT\_4} & \texttt{int}, \texttt{long}, \texttt{pointer}, \texttt{enum} and \texttt{struct} \\
\texttt{INT\_8} & \texttt{long long} \\
\texttt{REAL\_4} & \texttt{float} \\
\texttt{REAL\_8} & \texttt{double} and \texttt{long double} \\
\texttt{VOID} & \texttt{void} \\
\bottomrule
\end{tabular}
\caption{Relationship between the target variable (types grouped by size and representation) and the C high-level types.}
\label{table:type-relationships}
\end{table}

Table~\ref{table:type-relationships} shows the target variable defined for this first kind of models and their corresponding C type. For \texttt{INT\_2}, \texttt{INT\_8}, \texttt{REAL\_4} and \texttt{VOID}, the class used corresponds with a single high-level type. \texttt{REAL\_8} groups \texttt{double} and \texttt{long double}, while \texttt{INT\_8} considers all the C types returned in the 32-bit \texttt{eax} register (structs are actually returned as pointers).

\begin{figure}
\centering
\includegraphics*[angle=0,width=0.95\textwidth]{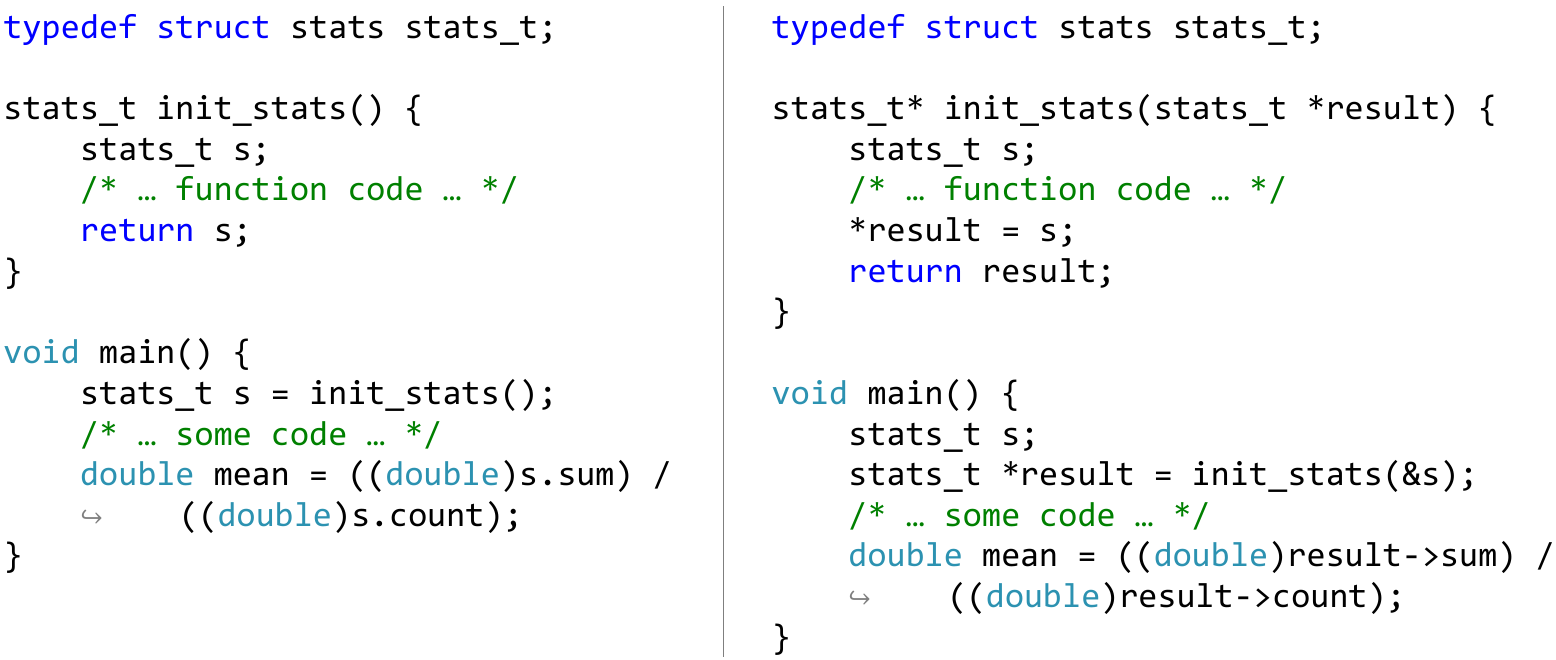}
\caption{The left-side code is transformed by \texttt{cl} into the right-side code to allow returning structs in \texttt{eax}, which are actually passed to the caller as pointers.}
\label{figure:code-transformation}
\end{figure}

Pointers are represented as \texttt{INT\_4} because the size of memory addresses in Intel x86 are 32 bits (4 bytes). The \texttt{struct} type is also clustered as \texttt{INT\_4}, because the \texttt{cl} compiler transforms returned structs into pointers to structs, as depicted in Figure~\ref{figure:code-transformation}. The returned pointer to \texttt{struct} is actually the \texttt{result} pointer passed as an argument. In this way, the actual \texttt{struct} is a local variable created in the scope of the caller (\texttt{s} variable in Figure~\ref{figure:code-transformation}), making easy the management of the memory allocated for the \texttt{struct}. This is the reason why the actual value returned is not a \texttt{struct} but a pointer (4~bytes).

The \texttt{union} type constructor is not listed in Table~\ref{table:type-relationships} because it has variable size and representation. When the size of the biggest field is not bigger than 32 bits, 4 bytes are used. When it is higher than 4 bytes and lower or equal to 8, 64 bits are used. In case it is greater than 8 bytes, the compiler generates the same code as for structs (Figure~\ref{figure:code-transformation}).

\subsubsection{Data size}
\label{subsubsection:data-size}

\begin{figure}[ht!]
\centering
\includegraphics*[angle=-90,width=\textwidth]{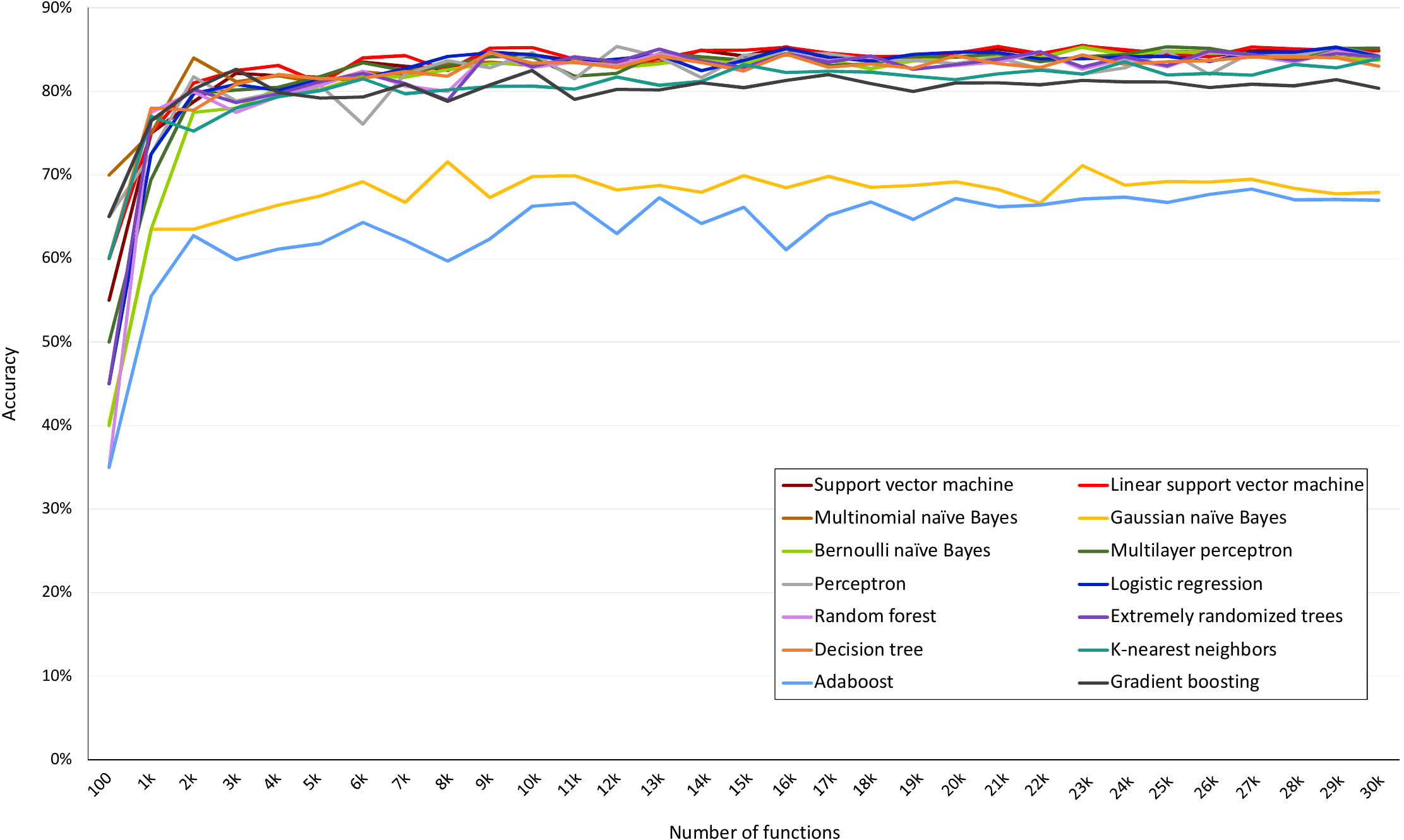}
\caption{Classifiers accuracy for increasing number of functions (classifiers of types with different size and representation).}
\label{figure:size-accuracy-size}
\end{figure}

\begin{figure}[ht!]
\centering
\includegraphics*[angle=-90,width=\textwidth]{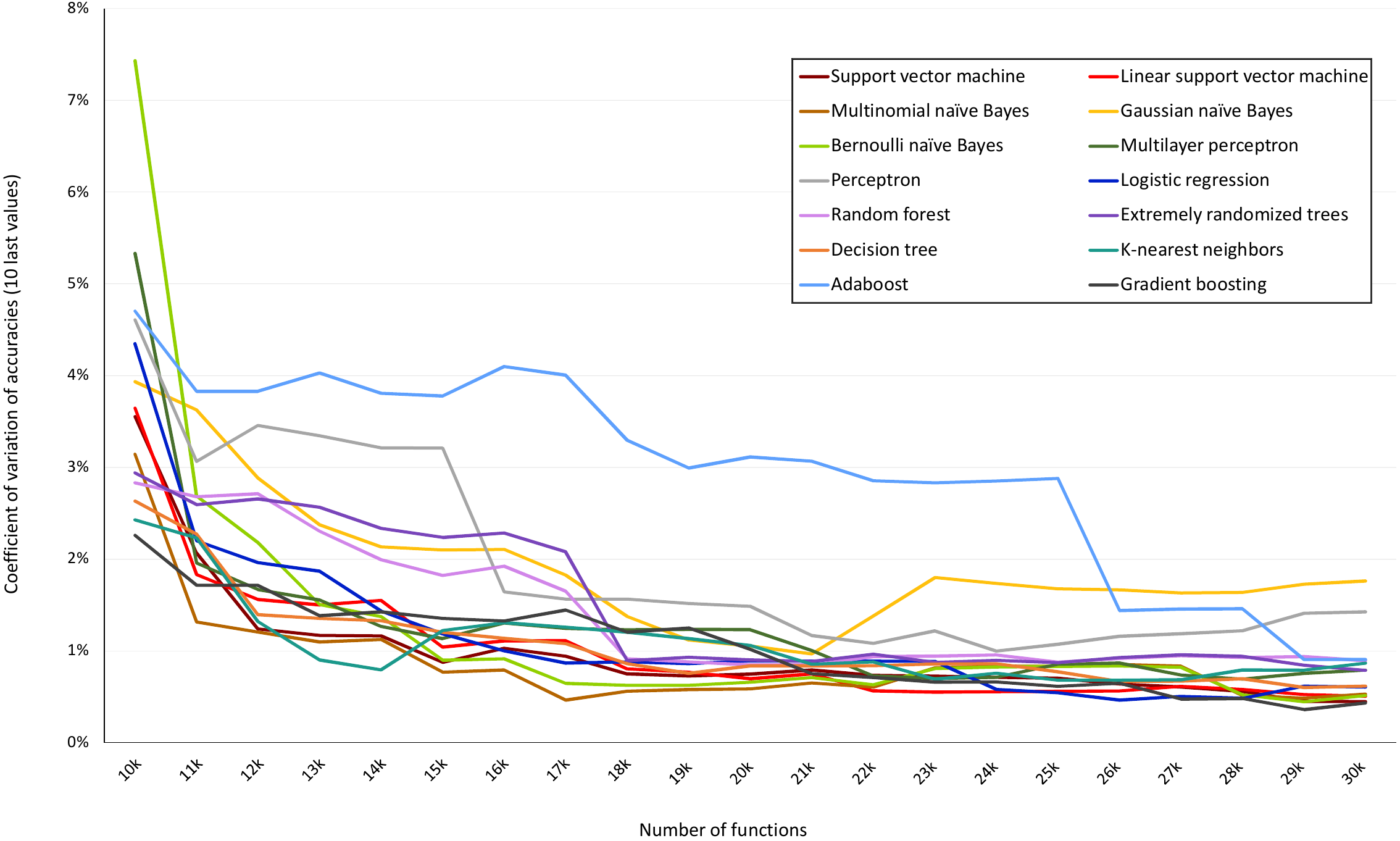}
\caption{Coefficient of variation of the last 10 accuracy values in Figure~\ref{figure:size-accuracy-size} (classifiers of types with different size and representation).}
\label{figure:size-cov-size}
\end{figure}

As mentioned in the methodology section, we use Cnerator to produce a dataset with such a number of functions that make models accuracies to converge. Figure~\ref{figure:size-accuracy-size} shows how classifiers accuracies grow as the dataset size increases. Figure~\ref{figure:size-cov-size} presents the CoV of the last 10 values. We can see how, with 26,000 functions, the CoVs of the accuracies for all the classifiers are below 2\%. Since CoV is computed for the last 10 iterations, and each iteration increases 1,000 functions, we build the dataset with 16,000 functions (and their invocations). In addition to those 16,000 instances, we add the 2,329 functions retrieved from real programs (Table~\ref{table:projects-used}).

\subsubsection{Feature selection}
\label{subsubsection:feature-selection}

\begin{table}
\centering
\scriptsize
\begin{tabular}{llr}
\toprule
Classifier & Applied method (wrapped algorithm, accuracy threshold) & Selected features \\
\midrule
AdaBoost & SelectFromModel(Random forest, Mean) & 134 \\
Bernoulli na\"ive Bayes & SelectFromModel(Random forest,  Median) & 439 \\
Decision tree & SelectFromModel(Extremely randomized trees, Median) & 419 \\
Extremely randomized trees & SelectFromModel(Extremely randomized trees, Median) & 419 \\
Gaussian na\"ive Bayes & SelectFromModel(Extremely randomized trees, Mean) & 130 \\
Gradient boosting & SelectFromModel(Extremely randomized trees, Median) & 419 \\
K-nearest neighbors & SelectFromModel(Random forest, Median) & 439 \\
Linear support vector machine & SelectFromModel(Random forest, Median) & 439 \\
Logistic regression & SelectFromModel(Random forest, Median) & 439 \\
Multilayer perceptron & SelectFromModel(Extremely randomized trees, Median) & 419 \\
Multinomial na\"ive Bayes & SelectFromModel(Random forest,  Median) & 439 \\
Perceptron & SelectFromModel(Random forest, Median) & 439 \\
Random forest & SelectFromModel(Random forest,  Mean) & 134 \\
Support vector machine & SelectFromModel(Extremely randomized trees, Median) & 419 \\
\bottomrule
\end{tabular}
\caption{Best feature selection method used for each classifier (classifiers of types with different size and representation).}
\label{table:size-feature-selection}
\end{table}

We apply the five feature-selection methods described in Section~\ref{subsection:feature-selection} (\texttt{RFECV} and \texttt{SelectFromModel} with random forest and extremely randomized trees, with mean and median thresholds) and select, for each classifier, the one with best accuracy. Table~\ref{table:size-feature-selection} shows the best feature selection method for each classifier, together with the number of features selected. The 1,019 features of the original dataset are reduced, on average, to 366. The selected features produce statistically significant higher accuracy for Gaussian na\"ive Bayes (3.22\% better) and multilayer perceptron (3.52\%). Moreover, the lower number of features reduces training times and overfitting.

\subsubsection{Hyperparameter tuning}
\label{subsubsection:hyperparameter-tuning}

We tune hyperparameters of the models as described in Section~\ref{subsection:hyperparameter-tuning}. For the hyperparameters found, AdaBoost increased its accuracy by 7.85\%. However, the rest of the classifiers obtained accuracy gains below 2\%, compared to the scikit-learn default parameters. The hyperparameters used for each model are detailed in~\cite{paperwebpage}.

\subsubsection{Results}
\label{subsubsection:results-grouped-types}

\begin{figure}
\centering
\includegraphics*[angle=-90,width=\textwidth]{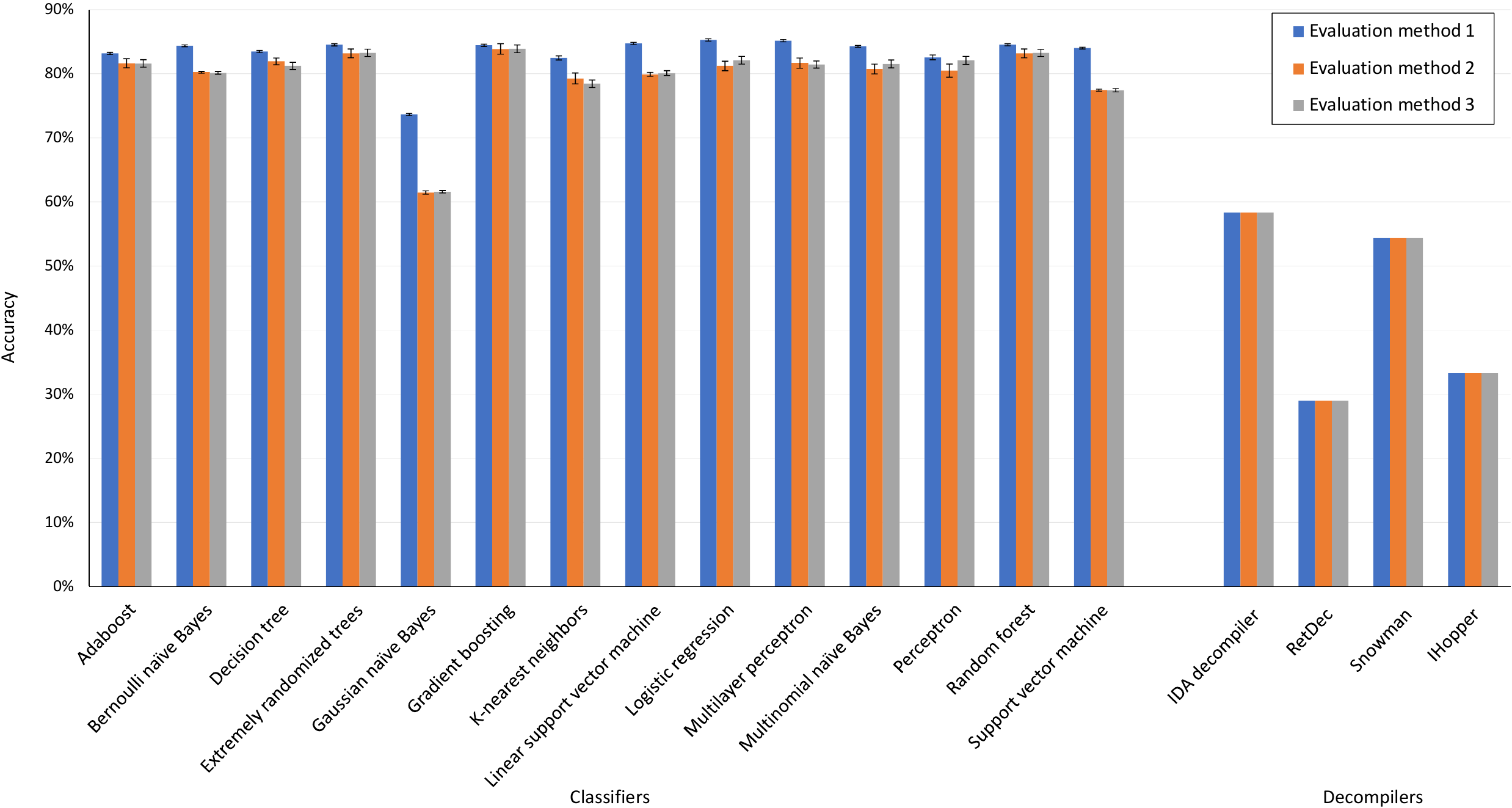}
\caption{Accuracies of our classifiers and the existing decompilers, using the three different evaluation methods described in Section~\ref{subsection:evaluation} (classifiers of types with different size and representation).}
\label{figure:size-accuracy-graph}
\end{figure}

Figure~\ref{figure:size-accuracy-graph} shows the accuracies of the 14 trained models (left-hand side) and the selected decompilers (right-hand side). All the systems are evaluated with the three methods described in Section~\ref{subsection:evaluation}. It can be seen how all the classifiers created with our dataset perform better than the existing decompilers, for all the evaluation methods.

In Figure~\ref{figure:size-accuracy-graph}, we can also see that there are significant differences between the first evaluation method and the two last ones, for all the machine learning models. This shows how the common evaluation method that takes 80\% for training and 20\% for testing is too optimistic for this project. We need to feed the models with sufficient code written by real programmers, so that we are able to predict return types with different programming styles. Existing decompilers show no influence on the evaluation method, because they use deterministic algorithms to infer return types.

\begin{figure}
\centering
\includegraphics*[angle=-90,width=\textwidth]{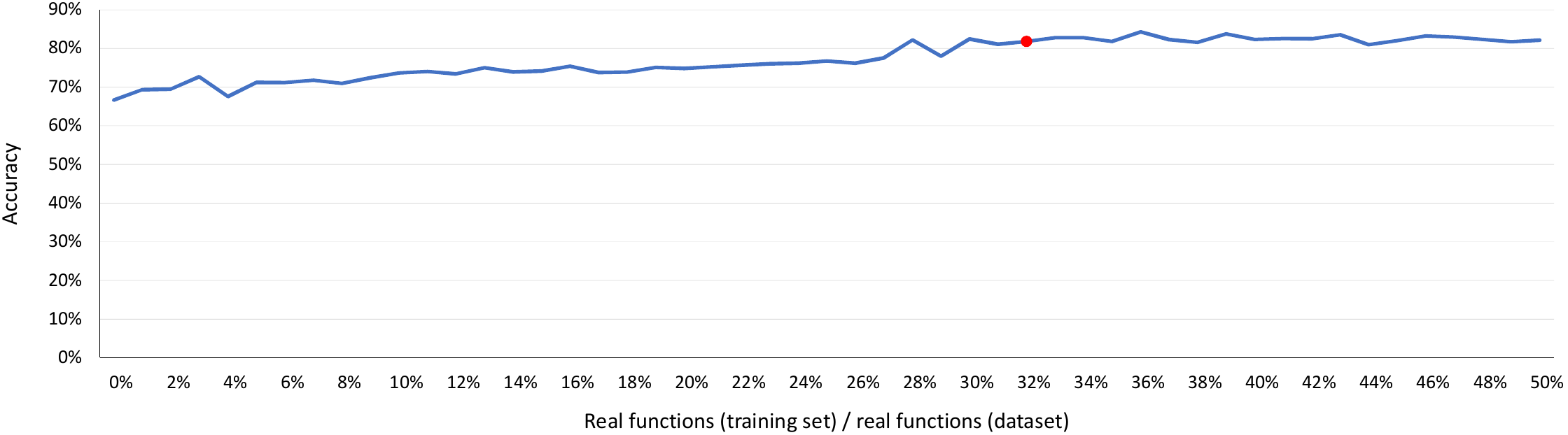}
\caption{Accuracy of a decision tree for different percentage of real functions included in the training dataset (classifiers of types with different size and representation). The red dot indicates the value where the CoV of the last 10 accuracies is lower than 1\%.}
\label{figure:size-r}
\end{figure}

\begin{table}
\centering
\scriptsize
\begin{tabular}{clr@{\hspace{1ex}}r@{\hspace{1ex}}rr@{\hspace{1ex}}r@{\hspace{1ex}}rr@{\hspace{1ex}}r@{\hspace{1ex}}rr@{\hspace{1ex}}r@{\hspace{1ex}}r}
\toprule
&  &\multicolumn{3}{c}{Accuracy}&\multicolumn{3}{c}{Precision}&\multicolumn{3}{c}{Recall}&\multicolumn{3}{c}{F\textsubscript{1}-score}\\
\midrule
\multirow{14}{*}{\begin{sideways}{Classifiers}\end{sideways}}
& AdaBoost & 0.816 & $\pm$ & 1.40\% & 0.821 & $\pm$ & 1.10\% & 0.815 & $\pm$ & 0.54\% & 0.805 & $\pm$ & 1.01\% \\
& Bernoulli na\"ive Bayes & 0.801 & $\pm$ & 0.51\% & 0.801 & $\pm$ & 0.43\% & 0.830 & $\pm$ & 0.35\% & 0.798 & $\pm$ & 0.42\% \\
& Decision tree & 0.812 & $\pm$ & 1.45\% & 0.817 & $\pm$ & 1.21\% & 0.814 & $\pm$ & 0.55\% & 0.803 & $\pm$ & 1.06\% \\
& Extremely randomized trees & \textbf{0.833} & $\pm$ & 1.39\% & 0.839 & $\pm$ & 1.22\% & \textbf{0.831} & $\pm$ & 0.50\% & 0.823 & $\pm$ & 1.04\% \\
& Gaussian na\"ive Bayes & 0.616 & $\pm$ & 0.59\% & 0.687 & $\pm$ & 0.61\% & 0.736 & $\pm$ & 0.35\% & 0.650 & $\pm$ & 0.60\% \\
& Gradient boosting & \textbf{0.839} & $\pm$ & 1.44\% &\textbf{ 0.863} & $\pm$ & 1.15\% & \textbf{0.832} & $\pm$ & 0.51\% & \textbf{0.838} & $\pm$ & 1.05\% \\
& K-nearest neighbors & 0.785 & $\pm$ & 1.51\% & 0.798 & $\pm$ & 1.22\% & 0.805 & $\pm$ & 0.54\% & 0.783 & $\pm$ & 1.07\% \\
& Linear support vector machine & 0.801 & $\pm$ & 0.95\% & 0.827 & $\pm$ & 1.32\% & 0.823 & $\pm$ & 0.43\% & 0.806 & $\pm$ & 0.90\% \\
& Logistic regression & 0.821 & $\pm$ & 1.52\% & 0.817 & $\pm$ & 1.37\% & \textbf{0.834} & $\pm$ & 0.50\% & 0.812 & $\pm$ & 1.15\% \\
& Multilayer perceptron & 0.814 & $\pm$ & 1.37\% & 0.821 & $\pm$ & 1.24\% & \textbf{0.831} & $\pm$ & 0.46\% & 0.809 & $\pm$ & 0.94\% \\
& Multinomial na\"ive Bayes & 0.815 & $\pm$ & 1.51\% & 0.819 & $\pm$ & 1.32\% & 0.829 & $\pm$ & 0.52\% & 0.807 & $\pm$ & 1.14\% \\
& Perceptron & 0.821 & $\pm$ & 1.53\% & 0.856 & $\pm$ & 1.74\% & 0.815 & $\pm$ & 0.82\% & 0.818 & $\pm$ & 1.28\% \\
& Random forest & \textbf{0.833} & $\pm$ & 1.37\% & 0.840 & $\pm$ & 1.15\% & \textbf{0.830} & $\pm$ & 0.49\% & 0.823 & $\pm$ & 1.00\% \\
& Support vector machine & 0.774 & $\pm$ & 0.64\% & 0.807 & $\pm$ & 1.35\% & 0.809 & $\pm$ & 0.37\% & 0.782 & $\pm$ & 0.70\% \\
\midrule
\multirow{4}{*}{\begin{sideways}{Decomp.}\end{sideways}}
& IDA decompiler & 0.583 & & & 0.495 & & & 0.413 & & & 0.415 \\
& RetDec & 0.290 & & & 0.111 & & & 0.133 & & & 0.110 \\
& Snowman & 0.544 & & & 0.365 & & & 0.328 & & & 0.322 \\
& Hopper & 0.333 & & & 0.132 & & & 0.132 & & & 0.079 \\
\bottomrule
\end{tabular}
\caption{Performance of the classifiers and existing decompilers using the third evaluation method (classifiers of types with different size and representation). 95\% confidence intervals are expressed as percentages. Bold font represents the best values. If one column has multiple cells in bold, it means that values are not significantly different.}
\label{table:size-performance}
\end{table}

One discussion related to the second evaluation method is finding out the number of real functions to be included in the training set, so that return types could be inferred for unknown real code. As described in Section~\ref{subsection:evaluation}, we start with a test set of 100\% synthesized functions, and incrementally add real functions until model accuracy converges. Figure~\ref{figure:size-r} shows this influence of real functions on the classifier accuracy. The red dot shows that with 32\% of real functions, CoV of model accuracy falls below 1\%. We fixed that value for the second evaluation method.

Figure~\ref{figure:size-accuracy-graph} also shows that there is no statistically significant difference between the second method and the third one (i.e., 95\% confidence intervals overlap~\cite{Georges2007}). This means that we need to include in the training dataset at least 32\% of real functions, so that the trained models are able to predict return types of code written by programmers not included in the training dataset.

Table~\ref{table:size-performance} shows the accuracy, precision, recall and F\textsubscript{1}-score of our models and the existing compilers, using the third evaluation method (Method~\ref{item:third} in Section~\ref{subsection:evaluation}). Gradient boosting is the classifier with the best results: 0.839 accuracy and 0.838 F\textsubscript{1}-score. Sometimes, there are no significant differences between random forest and extremely randomized trees. Gradient boosting accuracy and F\textsubscript{1}-score are, respectively, 43.9\% and 101.9\% higher than the best decompiler (IDA).

\subsection{Classifying with high-level types}
\label{subsection:high-level-types}

In the previous subsection, we measured the performance of our predictive models, considering types by their size and representation. However, the objective of a decompiler is to infer the high-level C types, even if they share the exact size and representation. Following the same methodology, we now conduct a new experiment to reconstruct C types from binary code.

In this case, we consider the C built-in types \texttt{bool}, \texttt{char}, \texttt{short}, \texttt{int}, \texttt{long long}, \texttt{float}, \texttt{double} and \texttt{void}. For the particular case of Microsoft \texttt{cl} and 32-bit architecture, the \texttt{long} type is exactly the same as \texttt{int}, since the semantic analyzer allows the very same operations and its target size and representation are the same; the same occurs for \texttt{double} and \texttt{long double}. For this reason, \texttt{long} and \texttt{long double} types are considered the same as, respectively, \texttt{int} and \texttt{double}. The \texttt{signed} and \texttt{unsigned} type specifiers are not considered, as there is no difference between their binary representations.

We also consider \texttt{pointer} and \texttt{struct} type constructors. Arrays are not classified because C functions cannot return arrays (they actually return pointers)~\cite{KernighanRitchie1978}. As discussed in Section~\ref{subsection:grouping-types}, the \texttt{union} type constructor is actually represented as one single variable with the size of the biggest field, so \texttt{cl} generates no different code when the type of the biggest field is used instead of \texttt{union}. The same happens with \texttt{enum} and \texttt{int}.

Our classifiers detect the \texttt{struct} and \texttt{pointer} type constructors, but not the \texttt{struct} fields or the pointed type. Once we know the return type is \texttt{pointer} or \texttt{struct}, the subtypes used to build the composite types could be obtained with existing deterministic approaches~\cite{Caballero2016, Mycroft1999}.

\begin{table}
\centering
\scriptsize
\begin{tabular}{ll|rrrrrrrrrr}
\toprule
& & \multicolumn{10}{c}{Predicted class} \\
& & \texttt{bool} & \texttt{char} & \texttt{short}  & \texttt{int}  & \texttt{pointer} & \texttt{struct} & \texttt{long long} & \texttt{float} & \texttt{double} & \texttt{void} \\
\midrule
\multirow{10}{*}{\begin{sideways}{Actual class}\end{sideways}}
& \texttt{bool} & \textbf{\textcolor{darkgreen}{483}} & \textbf{\textcolor{darkred}{61}} & 0 & 1 & 12 & 0 & 41 & 0 & 0 & 2 \\
& \texttt{char} & \textbf{\textcolor{darkred}{285}} & \textbf{\textcolor{darkgreen}{179}} & 60 & 16 & 13 & 0 & 45 & 0 & 0 & 2 \\
& \texttt{short} & 7 & 85 & 395 & 38 & 24 & 0 & 44 & 0 & 1 & 6 \\
& \texttt{int} & 4 & 38 & 67 & \textbf{\textcolor{darkgreen}{184}} & \textbf{\textcolor{darkred}{111}} & \textbf{\textcolor{darkred}{132}} & 59 & 0 & 0 & 5 \\
& \texttt{pointer} & 4 & 16 & 3 & \textbf{\textcolor{darkred}{48}} & \textbf{\textcolor{darkgreen}{365}} & \textbf{\textcolor{darkred}{106}} & 55 & 0 & 0 & 3 \\
& \texttt{struct} & 0 & 0 & 0 & \textbf{\textcolor{darkred}{10}} & \textbf{\textcolor{darkred}{6}} & \textbf{\textcolor{darkgreen}{584}} & 0 & 0 & 0 & 0 \\
& \texttt{long long} & 0 & 4 & 0 & 5 & 28 & 3 & 554 & 0 & 0 & 6 \\
& \texttt{float} & 0 & 1  & 1 & 0 & 16 & 0 & 40 & 358 & 184 & 0 \\
& \texttt{double} & 0 & 0 & 0 & 0 & 12 & 0 & 52 & 170 & 365 & 1 \\
& \texttt{void} & 3 & 2 & 0 & 2 & 3 & 0 & 41 & 0 & 0 & 549 \\
\bottomrule
\end{tabular}
\caption{Confusion matrix for the decision tree classification of high-level types, with a balanced dataset comprising 6,000 functions.}
\label{table:confusion-matrix-before}
\end{table}

We first analyze how well the exiting pattern generalizations (Table~\ref{table:basic-generalization-patterns}) classify high-level return types. To that end, we conduct the following experiment. First, we define the target variable as the high-level C types described above. Then, we build a decision tree classifier and evaluate it with a balanced test dataset comprising 6,000 functions. The confusion matrix obtained is shown in Table~\ref{table:confusion-matrix-before}. If we analyze the two types with 1-byte size (\texttt{bool} and \texttt{char}), we can see that 28.8\% of the instances are misclassified between \texttt{bool} and \texttt{char}. A similar misclassification issue occurs for the 4-byte-size types \texttt{int}, \texttt{struct} and \texttt{pointer}, mistaking 22.9\% of the instances among these three types. 

\begingroup
\centering
\scriptsize
\begin{longtable}{ll}
\toprule
Assembly pattern & Feature (Generalization) \\
\midrule
\endfirsthead
\toprule
Assembly pattern & Feature (Generalization) \\
\midrule
\endhead
\multicolumn{2}{c}{(continues)}\\
\midrule
\endfoot
\bottomrule
\\
\caption{New generalizations added to improve the classification of high-level return types. \textit{reg} variables represent registers, \textit{literal} integer literals, \textit{address} absolute addresses, \textit{*address} absolute addresses dereferences and \textit{offset} relative addresses. Variables between normal-font brackets ([]) represent optional arguments, while typewriter-font brackets (\texttt{[]}) are the assembly brackets denoting register-based dereferences.}
\label{table:advanced-generalization-patterns}
\endlastfoot

\textit{cond\_jmp} \textit{offset\textsubscript{1}} & \multirow{11}{*}{\textit{bool\_cast}(\textit{reg\textsubscript{1}})} \\*
\texttt{mov} \texttt{[}\textit{reg\textsubscript{1}}\texttt{]}, \textit{literal\textsubscript{1}} & \\*
\texttt{jmp} \textit{offset\textsubscript{2}} & \\*
\texttt{mov} \texttt{[}\textit{reg\textsubscript{1}}\texttt{]}, \textit{literal\textsubscript{2}} & \\*
 & \\*
where: \textit{cond\_jmp} $\in$ \{\texttt{jo}, \texttt{jno}, \texttt{js}, \texttt{jns}, \texttt{je}, \texttt{jz}, \texttt{jne}, \texttt{jnz}, \texttt{jb}, \texttt{jnae}, \texttt{jc}, & \\*
\hspace{8ex}{\tiny$\hookrightarrow$}\hspace{2ex}\texttt{jnb}, \texttt{jae}, \texttt{jnc}, \texttt{jbe}, \texttt{jna}, \texttt{ja}, \texttt{jnbe}, \texttt{jl}, \texttt{jnge}, \texttt{jge}, \texttt{jnl}, \texttt{jle}, & \\*
\hspace{8ex}{\tiny$\hookrightarrow$}\hspace{2ex}\texttt{jng}, \texttt{jg}, \texttt{jnle}, \texttt{jp}, \texttt{jpe}, \texttt{jnp}, \texttt{jpo}, \texttt{jcxz}, \texttt{jecxz}\} & \\*
\hspace{8ex}\textit{literal\textsubscript{1}}, \textit{literal\textsubscript{2}} $\in$ \{\texttt{0}, \texttt{1}\} & \\*
\hspace{8ex}\textit{literal\textsubscript{1}} $\neq$ \textit{literal\textsubscript{2}} & \\*
\hspace{8ex}\textit{offset\textsubscript{1}} $\neq$ \textit{offset\textsubscript{2}} & \\

\midrule

\textit{mov} \textit{arg\textsubscript{2}}, \textit{arg\textsubscript{1}} & \multirow{8}{*}{\textit{gen\_mov\_chain}(\textit{arg\textsubscript{n}}, ..., \textit{arg\textsubscript{1}})} \\*
\textit{mov} \textit{arg\textsubscript{3}}, \textit{arg\textsubscript{2}} & \\*
... & \\*
\textit{mov} \textit{arg\textsubscript{n}}, \textit{arg\textsubscript{n-1}} & \\
 & \\*
where: \textit{mov} $\in$ \{\texttt{mov}, \texttt{movzx}, \texttt{movsx}\} & \\*
\hspace{8ex}\textit{arg\textsubscript{1}} $\in$ \{\textit{reg}, \texttt{[}\textit{reg}\texttt{]}, \textit{*address}, \textit{literal}\} & \\*
\hspace{8ex}\textit{arg\textsubscript{2}}, ..., \textit{arg\textsubscript{n}} $\in$ \{\textit{reg}, \texttt{[}\textit{reg}\texttt{]}, \textit{*address}\} & \\

\midrule

\texttt{mov} \textit{arg\textsubscript{2}}, \textit{arg\textsubscript{1}} & \multirow{7}{*}{\textit{mov\_chain}(\textit{arg\textsubscript{n}}, ..., \textit{arg\textsubscript{1}})} \\*
\texttt{mov} \textit{arg\textsubscript{3}}, \textit{arg\textsubscript{2}} & \\*
... & \\*
\texttt{mov} \textit{arg\textsubscript{n}}, \textit{arg\textsubscript{n-1}} & \\*
 & \\*
where: \textit{arg\textsubscript{1}} $\in$ \{\textit{reg}, \texttt{[}\textit{reg}\texttt{]}, \textit{*address}, \textit{literal}\} & \\*
\hspace{8ex}\textit{arg\textsubscript{2}}, ..., \textit{arg\textsubscript{n}} $\in$ \{\textit{reg}, \texttt{[}\textit{reg}\texttt{]}, \textit{*address}\} & \\

\midrule

\textit{bool\_cast}(\textit{reg\textsubscript{1}}) & \multirow{5}{*}{\textit{return\_bool\_cast}(\textit{reg\textsubscript{ax}}, \textit{reg\textsubscript{n}}, ..., \textit{reg\textsubscript{1}})} \\*
\textit{mov\_chain}(\textit{reg\textsubscript{ax}}, \textit{reg\textsubscript{n}}, ...,  \textit{reg\textsubscript{1}}) \\*
\textit{callee\_epilogue} & \\*
 & \\*
where: \textit{reg\textsubscript{ax}} $\in$ \{\texttt{eax}, \texttt{ax}, \texttt{al}, \texttt{ah}\} & \\

\midrule

\textit{mov\_chain}(\textit{reg\textsubscript{ax}}, \textit{reg\textsubscript{n}}, ..., \textit{reg\textsubscript{1}}, \textit{literal}) & \multirow{5}{*}{\textit{return\_bool\_literal}(\textit{reg\textsubscript{ax}}, \textit{reg\textsubscript{n}}, ..., \textit{reg\textsubscript{1}})} \\*
\textit{callee\_epilogue} & \\*
& \\*
where: \textit{reg\textsubscript{ax}} $\in$ \{\texttt{eax}, \texttt{ax}, \texttt{al}, \texttt{ah}\} & \\*
\hspace{8ex}\textit{literal} $\in$ \{\texttt{0}, \texttt{1}\} & \\

\midrule

\textit{gen\_mov\_chain}(\textit{reg\textsubscript{2}}, \textit{reg\textsubscript{1}}) & \multirow{3}{*}{\textit{inc\_int}(\textit{reg\textsubscript{1}})} \\*
\texttt{add} \textit{reg\textsubscript{2}}, \texttt{1} & \\*
\textit{mov\_chain}(\textit{reg\textsubscript{1}}, \textit{reg\textsubscript{2}})& \\

\midrule

\textit{gen\_mov\_chain}(\textit{reg\textsubscript{2}}, \textit{reg\textsubscript{1}}) & \multirow{3}{*}{\textit{dec\_int}(\textit{reg\textsubscript{1}})} \\*
\texttt{sub} \textit{reg\textsubscript{2}}, \texttt{1} & \\*
\textit{mov\_chain}(\textit{reg\textsubscript{1}}, \textit{reg\textsubscript{2}})& \\

\midrule

\textit{gen\_mov\_chain}(\textit{reg\textsubscript{2}}, \textit{reg\textsubscript{1}}) & \multirow{5}{*}{\textit{inc\_ptr}(\textit{reg\textsubscript{1}})} \\*
\texttt{add} \textit{reg\textsubscript{2}}, \textit{literal} & \\*
\textit{mov\_chain}(\textit{reg\textsubscript{1}}, \textit{reg\textsubscript{2}})& \\*
& \\*
where: \textit{literal} $>$ 1 & \\

\midrule

\textit{gen\_mov\_chain}(\textit{reg\textsubscript{2}}, \textit{reg\textsubscript{1}}) & \multirow{5}{*}{\textit{dec\_ptr}(\textit{reg\textsubscript{1}})} \\*
\texttt{sub} \textit{reg\textsubscript{2}}, \textit{literal} & \\*
\textit{mov\_chain}(\textit{reg\textsubscript{1}}, \textit{reg\textsubscript{2}})& \\*
& \\*
where: \textit{literal} $>$ 1 & \\

\midrule

\texttt{shl} \textit{reg\textsubscript{2}}, \textit{literal} & \multirow{3}{*}{\textit{sub\_int\_to\_ptr}(\textit{reg\textsubscript{1}}, \textit{*address}, \textit{reg\textsubscript{2}})} \\*
\textit{mov\_chain}(\textit{reg\textsubscript{1}}, \textit{*address}) & \\*
\texttt{sub} \textit{reg\textsubscript{1}}, \textit{reg\textsubscript{2}} & \\

\midrule

\texttt{call} \texttt{\_\_allmul} & \multirow{3}{*}{\textit{sub\_int\_to\_ptr}(\textit{reg}, \textit{*address})} \\*
\textit{mov\_chain}(\textit{reg}, \textit{*address}) & \\*
\texttt{sub} \textit{reg}, \texttt{eax} & \\

\midrule

\texttt{call} \texttt{\_\_allmul} & \multirow{2}{*}{\textit{add\_int\_to\_ptr}(\textit{*address})} \\*
\texttt{add} \texttt{eax}, \textit{*address} & \\

\midrule

\textit{sub\_int\_to\_ptr}(\textit{reg\textsubscript{1}}, \textit{*address}, [\textit{reg\textsubscript{2}}]) & \multirow{5}{*}{\shortstack[l]{\textit{return\_sub\_int\_to\_ptr}(\textit{reg\textsubscript{ax}}, \textit{reg\textsubscript{1}}, \textit{*address}, \\ {\tiny$\hookrightarrow$}\hspace{2ex}[\textit{reg\textsubscript{2}}])}} \\*
\textit{mov\_chain}(\textit{reg\textsubscript{ax}}, \textit{reg\textsubscript{1}})& \\*
\textit{callee\_epilogue} & \\*
& \\*
where: \textit{reg\textsubscript{ax}} $\in$ \{\texttt{eax}, \texttt{ax}, \texttt{al}, \texttt{ah}\} & \\

\midrule

\textit{sub\_int\_to\_ptr}(\textit{reg\textsubscript{1}}, \textit{*address}, [\textit{reg\textsubscript{2}}]) & \multirow{6}{*}{\shortstack[l]{\textit{return\_sub\_assign\_int\_to\_ptr}(\textit{reg\textsubscript{ax}}, \textit{arg\textsubscript{1}},\\ {\tiny$\hookrightarrow$}\hspace{2ex}\textit{reg\textsubscript{1}}, \textit{*address}, [\textit{reg\textsubscript{2}}])}} \\*
\textit{gen\_mov\_chain}(\textit{reg\textsubscript{ax}}, \textit{arg\textsubscript{1}}, \textit{reg\textsubscript{1}}) & \\*
\textit{callee\_epilogue} & \\*
& \\*
where:  \textit{reg\textsubscript{ax}} $\in$ \{\texttt{eax}, \texttt{ax}, \texttt{al}, \texttt{ah}\} & \\*
\hspace{8ex}\textit{arg\textsubscript{1}} $\in$ \{\textit{reg}, \texttt{[}\textit{reg}\texttt{]}, \textit{*address}\} & \\

\midrule

\textit{add\_int\_to\_ptr}(\textit{reg\textsubscript{1}}, \textit{*address}) & \multirow{6}{*}{\shortstack[l]{\textit{return\_add\_assign\_int\_to\_ptr}(\textit{reg\textsubscript{ax}}, \textit{arg\textsubscript{1}},\\ {\tiny$\hookrightarrow$}\hspace{2ex}\textit{reg\textsubscript{1}}, \textit{*address})}} \\*
\textit{gen\_mov\_chain}(\textit{reg\textsubscript{ax}}, \textit{arg\textsubscript{1}}, \textit{reg\textsubscript{1}}) & \\*
\textit{callee\_epilogue} & \\*
& \\*
where:  \textit{reg\textsubscript{ax}} $\in$ \{\texttt{eax}, \texttt{ax}, \texttt{al}, \texttt{ah}\} & \\*
\hspace{8ex}\textit{arg\textsubscript{1}} $\in$ \{\textit{reg}, \texttt{[}\textit{reg}\texttt{]}, \textit{*address}\} & \\

\midrule

\texttt{lea} \textit{reg\textsubscript{1}}, \textit{*address} & \multirow{5}{*}{\shortstack[l]{\textit{return\_add\_assign\_int\_to\_ptr}(\textit{reg\textsubscript{ax}}, \textit{arg\textsubscript{1}},\\ {\tiny$\hookrightarrow$}\hspace{2ex}\textit{reg\textsubscript{1}}, \textit{*address})}} \\*
\textit{gen\_mov\_chain}(\textit{reg\textsubscript{ax}}, \textit{arg\textsubscript{1}}, \textit{reg\textsubscript{1}}) & \\*
\textit{callee\_epilogue} & \\*
& \\*
where:  \textit{reg\textsubscript{ax}} $\in$ \{\texttt{eax}, \texttt{ax}, \texttt{al}, \texttt{ah}\} & \\*
\hspace{8ex}\textit{arg\textsubscript{1}} $\in$ \{\textit{reg}, \texttt{[}\textit{reg}\texttt{]}, \textit{*address}\} & \\

\midrule

\textit{math\_op} \textit{arg\textsubscript{1}} & \multirow{5}{*}{\textit{return\_int\_math\_op}(\textit{arg\textsubscript{1}})} \\*
\textit{callee\_epilogue} & \\*
& \\*
where: \textit{math\_op} $\in$ \{\texttt{div}, \texttt{idiv}\} & \\*
\hspace{8ex}\textit{arg\textsubscript{1}} $\in$ \{\textit{reg}, \texttt{[}\textit{reg}\texttt{]}, \textit{*address}\} & \\

\midrule

\textit{math\_op} \textit{arg\textsubscript{1}} & \multirow{8}{*}{\textit{return\_int\_math\_op}(\textit{reg\textsubscript{ax}}, \textit{arg\textsubscript{1}})} \\*
\textit{mov\_chain}(\textit{reg\textsubscript{ax}}, \textit{reg\textsubscript{rem}}) & \\*
\textit{callee\_epilogue} & \\*
& \\*
where: \textit{math\_op} $\in$ \{\texttt{div}, \texttt{idiv}\} & \\*
\hspace{8ex}\textit{arg\textsubscript{1}} $\in$ \{\textit{reg}, \texttt{[}\textit{reg}\texttt{]}, \textit{*address}\} & \\*
\hspace{8ex}\textit{reg\textsubscript{ax}} $\in$ \{\texttt{eax}, \texttt{ax}, \texttt{al}\} & \\*
\hspace{8ex}\textit{reg\textsubscript{rem}} $\in$ \{\texttt{edx}, \texttt{dx}, \texttt{ah}\} & \\

\midrule

\textit{math\_op} \textit{arg\textsubscript{1}} \textit{arg\textsubscript{2}} & \multirow{9}{*}{\textit{return\_int\_math\_op}(\textit{reg\textsubscript{ax}}, \textit{arg\textsubscript{1}}, \textit{arg\textsubscript{2}})} \\*
\textit{mov\_chain}(\textit{arg\textsubscript{ax}}, \textit{reg\textsubscript{1}}) & \\*
\textit{callee\_epilogue} & \\*
& \\*
where: \textit{math\_op} $\in$ \{\texttt{imul}, \texttt{sub}, \texttt{add}, \texttt{sar}, \texttt{sal}, \texttt{shr}, \texttt{shl}, \texttt{xor}, \texttt{or}, \texttt{and}\} & \\*
\hspace{8ex}\textit{arg\textsubscript{1}} $\in$ \{\textit{reg}, \texttt{[}\textit{reg}\texttt{]}, \textit{*address}\} & \\*
\hspace{8ex}\textit{arg\textsubscript{2}} $\in$ \{\textit{reg}, \texttt{[}\textit{reg}\texttt{]}, \textit{*address}, \textit{literal}\} & \\*
\hspace{8ex}\textit{reg\textsubscript{ax}} $\in$ \{\texttt{eax}, \texttt{ax}, \texttt{al}, \texttt{ah}\} & \\

\midrule

\textit{math\_op} \textit{arg\textsubscript{2}} & \multirow{7}{*}{\textit{return\_int\_math\_op\_assign}(\textit{reg\textsubscript{ax}}, \textit{arg\textsubscript{1}}, \textit{arg\textsubscript{2}})} \\*
\textit{mov\_chain}(\textit{reg\textsubscript{ax}}, \textit{arg\textsubscript{1}}) & \\*
\textit{callee\_epilogue} & \\*
& \\*
where: \textit{math\_op} $\in$ \{\texttt{div}, \texttt{idiv}\} & \\*
\hspace{8ex}\textit{arg\textsubscript{1}} $\in$ \{\textit{reg}, \texttt{[}\textit{reg}\texttt{]}, \textit{*address}\} & \\*
\hspace{8ex}\textit{arg\textsubscript{2}} $\in$ \{\textit{reg}, \texttt{[}\textit{reg}\texttt{]}, \textit{*address}\} & \\*
\hspace{8ex}\textit{reg\textsubscript{ax}} $\in$ \{\texttt{eax}, \texttt{ax}, \texttt{al}, \texttt{ah}\} & \\

\midrule

\textit{math\_op} \textit{arg\textsubscript{2}} & \multirow{7}{*}{\textit{return\_int\_math\_op\_assign}(\textit{reg\textsubscript{ax}}, \textit{arg\textsubscript{1}}, \textit{arg\textsubscript{2}})} \\*
\textit{mov\_chain}(\textit{reg\textsubscript{ax}}, \textit{arg\textsubscript{1}}, \textit{reg\textsubscript{rem}}) & \\*
\textit{callee\_epilogue} & \\*
& \\*
where: \textit{math\_op} $\in$ \{\texttt{div}, \texttt{idiv}\} & \\*
\hspace{8ex}\textit{arg\textsubscript{1}} $\in$ \{\textit{reg}, \texttt{[}\textit{reg}\texttt{]}, \textit{*address}\} & \\*
\hspace{8ex}\textit{arg\textsubscript{2}} $\in$ \{\textit{reg}, \texttt{[}\textit{reg}\texttt{]}, \textit{*address}\} & \\*
\hspace{8ex}\textit{reg\textsubscript{rem}} $\in$ \{\texttt{edx}, \texttt{dx}, \texttt{ah}\} & \\

\midrule

\textit{math\_op} \textit{arg\textsubscript{2}} \textit{arg\textsubscript{3}} & \multirow{10}{*}{\shortstack[l]{\textit{return\_int\_math\_op\_assign}(\textit{reg\textsubscript{ax}}, \textit{arg\textsubscript{1}}, \textit{arg\textsubscript{2}},\\ {\tiny$\hookrightarrow$}\hspace{2ex}\textit{arg\textsubscript{3}})}} \\*
\textit{mov\_chain}(\textit{reg\textsubscript{ax}}, \textit{arg\textsubscript{1}}, \textit{arg\textsubscript{2}}) & \\*
\textit{callee\_epilogue} & \\*
& \\*
where: \textit{math\_op} $\in$ \{\texttt{imul}, \texttt{sub}, \texttt{add}, \texttt{sar}, \texttt{sal}, \texttt{shr}, \texttt{shl}, \texttt{xor}, \texttt{or}, \texttt{and}\} & \\*
\hspace{8ex}\textit{arg\textsubscript{1}} $\in$ \{\textit{reg}, \texttt{[}\textit{reg}\texttt{]}, \textit{*address}\} & \\*
\hspace{8ex}\textit{arg\textsubscript{2}} $\in$ \{\textit{reg}, \texttt{[}\textit{reg}\texttt{]}, \textit{*address}\} & \\*
\hspace{8ex}\textit{arg\textsubscript{3}} $\in$ \{\textit{reg}, \texttt{[}\textit{reg}\texttt{]}, \textit{*address}, \textit{literal}\} & \\*
\hspace{8ex}\textit{reg\textsubscript{ax}} $\in$ \{\texttt{eax}, \texttt{ax}, \texttt{al}, \texttt{ah}\} & \\

\end{longtable}
\endgroup

This experiment shows us how there is still room for improving the classification of high-level types with similar size and representation. Aware of that, we include new generalization patterns aimed at differentiating among high-level types with the same size and representation. To this end, we search for the misclassified functions in Table~\ref{table:confusion-matrix-before} and analyze the sequences of assembly code related to the same type. Once detected, we generalize and include them in our pattern extractor (Figure~\ref{figure:system-architecture}) to improve the classifiers. Therefore, we use machine learning to detect some of the limitations of the existing classifiers, analyze potential binary patterns, define new features of the dataset, and create better models to classify the return type of decompiled functions. The new generalizations defined are detailed in Table~\ref{table:advanced-generalization-patterns}.

\begin{figure}
\centering
\includegraphics*[angle=-90,width=\textwidth]{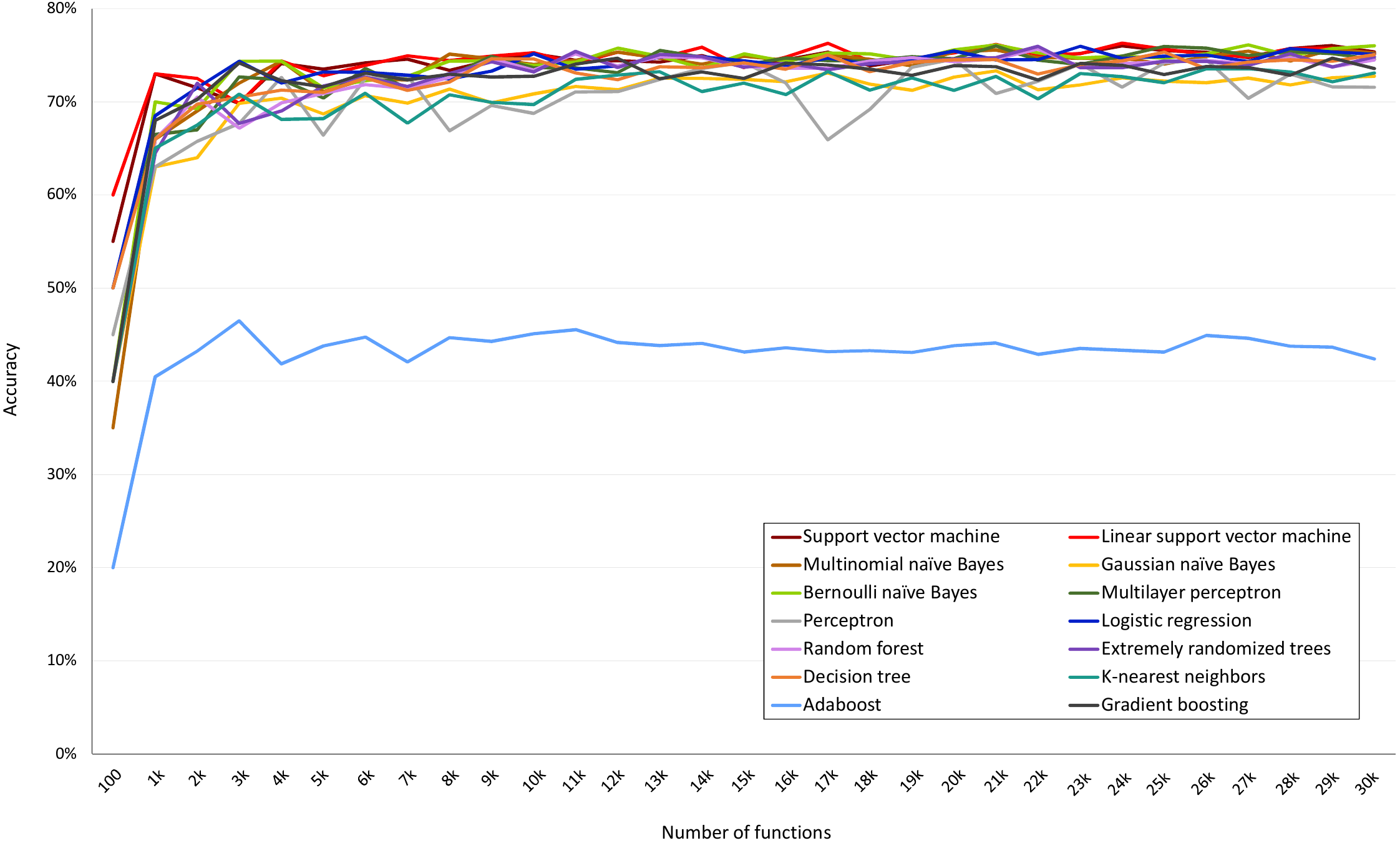}
\caption{Classifiers accuracy for increasing number of functions (classifiers of high-level types).}
\label{figure:all-accuracy-size}
\end{figure}

\begin{table}
\centering
\scriptsize
\begin{tabular}{llr}
\toprule
Classifier & Applied method (wrapped algorithm, accuracy threshold) & Selected features \\
\midrule
AdaBoost & SelectFromModel(Random forest, Mean) & 147 \\
Bernoulli na\"ive Bayes & SelectFromModel(Extremely randomized trees,  Median) & 456 \\
Decision tree & SelectFromModel(Random forest, Median) & 455 \\
Extremely randomized trees & SelectFromModel(Random forest, Median) & 455 \\
Gaussian na\"ive Bayes & SelectFromModel(Extremely randomized trees, Median) & 456 \\
Gradient boosting & SelectFromModel(Extremely randomized trees, Median) & 456 \\
K-nearest neighbors & SelectFromModel(Random forest, Median) & 455 \\
Linear support vector machine & SelectFromModel(Random forest, Median) & 455 \\
Logistic regression & SelectFromModel(Random forest, Median) & 455 \\
Multilayer perceptron & SelectFromModel(Extremely randomized trees, Median) & 456 \\
Multinomial na\"ive Bayes & SelectFromModel(Extremely randomized trees,  Median) & 456 \\
Perceptron & SelectFromModel(Random forest, Median) & 455 \\
Random forest & SelectFromModel(Random forest,  Median) & 455 \\
Support vector machine & SelectFromModel(Random forest,  Median) & 455 \\
\bottomrule
\end{tabular}
\caption{Best feature selection method used for each classifier (classifiers of high-level types).}
\label{table:all-feature-selection}
\end{table}

A new dataset is created with all the new generalization features in Table~\ref{table:advanced-generalization-patterns} to classify the ten different high-level types mentioned. We compute the optimal size of the dataset with the algorithm described in Section~\ref{subsection:dataset-size}. The resulting dataset has 18,000 synthetic functions (Figure~\ref{figure:all-accuracy-size}) plus the 2,339 functions implemented by real programmers. Following the methodology described in Section~\ref{section:methodology}, we run different feature selection algorithms, obtaining the features in Table~\ref{table:all-feature-selection}. The existing 1,036 features were reduced, on average, to 433. We then tune the hyperparameters of the models, achieving similar values to the previous experiment (their values can be consulted in~\cite{paperwebpage}).

\subsubsection{Results}
\label{subsubsection:results-high-level-types}

\begin{figure}
\centering
\includegraphics*[angle=-90,width=\textwidth]{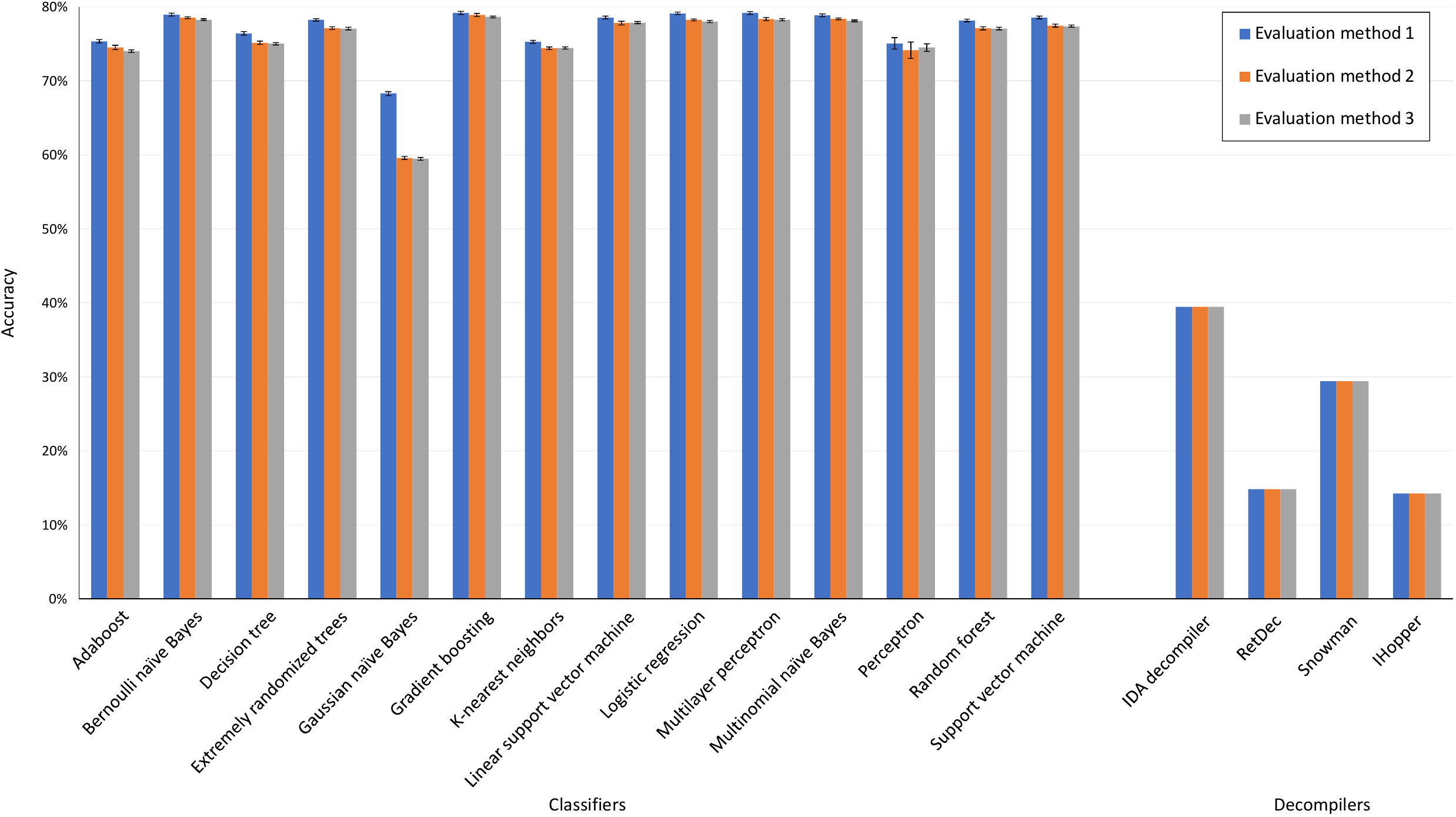}
\caption{Accuracies of our classifiers and the existing decompilers, using the three different evaluation methods described in Section~\ref{subsection:evaluation} (classifiers of high-level types).}
\label{figure:all-accuracy-graph}
\end{figure}

\begin{table}
\centering
\scriptsize
\begin{tabular}{clr@{\hspace{1ex}}r@{\hspace{1ex}}rr@{\hspace{1ex}}r@{\hspace{1ex}}rr@{\hspace{1ex}}r@{\hspace{1ex}}rr@{\hspace{1ex}}r@{\hspace{1ex}}r}
\toprule
&  &\multicolumn{3}{c}{Accuracy}&\multicolumn{3}{c}{Precision}&\multicolumn{3}{c}{Recall}&\multicolumn{3}{c}{F\textsubscript{1}-score}\\
\midrule
\multirow{14}{*}{\begin{sideways}{Classifiers}\end{sideways}}
& AdaBoost & 0.740 & $\pm$ & 0.49\% & 0.764 & $\pm$ & 0.92\% & 0.741 & $\pm$ & 0.42\% & 0.741 & $\pm$ & 0.65\% \\
& Bernoulli na\"ive Bayes & 0.783 & $\pm$ & 0.29\% & \textbf{0.822} & $\pm$ & 0.33\% & \textbf{0.785} & $\pm$ & 0.31\% & \textbf{0.792} & $\pm$ & 0.31\% \\
& Decision tree & 0.750 & $\pm$ & 0.47\% & 0.772 & $\pm$ & 0.85\% & 0.753 & $\pm$ & 0.47\% & 0.751 & $\pm$ & 0.59\% \\
& Extremely randomized trees & 0.771 & $\pm$ & 0.42\% & 0.793 & $\pm$ & 0.82\% & 0.774 & $\pm$ & 0.45\% & 0.772 & $\pm$ & 0.56\% \\
& Gaussian na\"ive Bayes & 0.595 & $\pm$ & 0.67\% & 0.671 & $\pm$ & 0.88\% & 0.692 & $\pm$ & 0.35\% & 0.619 & $\pm$ & 0.51\% \\
& Gradient boosting & \textbf{0.786} & $\pm$ & 0.31\% & \textbf{0.820} & $\pm$ & 0.29\% & 0.783 & $\pm$ & 0.31\% & \textbf{0.791} & $\pm$ & 0.30\% \\
& K-nearest neighbors & 0.745 & $\pm$ & 0.40\% & 0.759 & $\pm$ & 0.78\% & 0.749 & $\pm$ & 0.41\% & 0.743 & $\pm$ & 0.57\% \\
& Linear support vector machine & 0.779 & $\pm$ & 0.42\% & 0.803 & $\pm$ & 0.81\% & 0.782 & $\pm$ & 0.34\% & 0.780 & $\pm$ & 0.59\% \\
& Logistic regression & 0.780 & $\pm$ & 0.36\% & 0.802 & $\pm$ & 0.58\% & \textbf{0.785} & $\pm$ & 0.36\% & 0.785 & $\pm$ & 0.37\% \\
& Multilayer perceptron & 0.783 & $\pm$ & 0.36\% & 0.815 & $\pm$ & 0.65\% & \textbf{0.784} & $\pm$ & 0.37\% & 0.788 & $\pm$ & 0.43\% \\
& Multinomial na\"ive Bayes & 0.781 & $\pm$ & 0.28\% & 0.818 & $\pm$ & 0.29\% & \textbf{0.785} & $\pm$ & 0.31\% & 0.789 & $\pm$ & 0.30\% \\
& Perceptron & 0.745 & $\pm$ & 1.35\% & 0.789 & $\pm$ & 1.20\% & 0.747 & $\pm$ & 1.18\% & 0.749 & $\pm$ & 1.23\% \\
& Random forest & 0.771 & $\pm$ & 0.43\% & 0.796 & $\pm$ & 0.75\% & 0.773 & $\pm$ & 0.46\% & 0.773 & $\pm$ & 0.52\% \\
& Support vector machine & 0.774 & $\pm$ & 0.35\% & 0.816 & $\pm$ & 0.34\% & 0.779 & $\pm$ & 0.33\% & 0.784 & $\pm$ & 0.32\% \\
\midrule
\multirow{4}{*}{\begin{sideways}{Decomp.}\end{sideways}}
& IDA decompiler & 0.40 & & & 0.33 & & & 0.34 & & & 0.30 \\
& RetDec & 0.15 & & & 0.06 & & & 0.10 & & & 0.06 \\
& Snowman & 0.29 & & & 0.22 & & & 0.26 & & & 0.21 \\
& Hopper & 0.14 & & & 0.08 & & & 0.09 & & & 0.03 \\
\bottomrule
\end{tabular}
\caption{Performance of the classifiers and existing decompilers using the third evaluation method (classifiers of types with different size and representation). 95\% confidence intervals are expressed as percentages. Bold font represents the best values. If one column has multiple cells in bold, it means that values are not significantly different.}
\label{table:all-performance}
\end{table}

\begin{figure}
\centering
\includegraphics*[angle=-90,width=\textwidth]{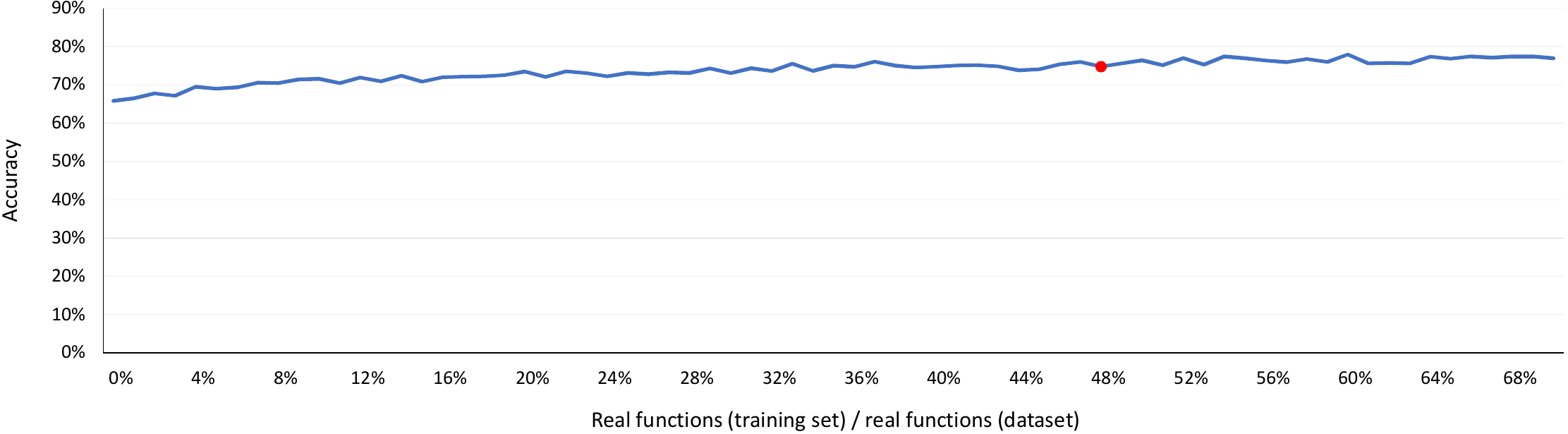}
\caption{Accuracy of a decision tree for different percentage of real functions included in the training dataset (classifiers of high-level types). The red dot indicates the value where the CoV of the last 10 accuracies is lower than 1\%.}
\label{figure:all-r}
\end{figure}

Figure~\ref{figure:all-accuracy-graph} compares the accuracies of the new models and the selected decompilers (detailed data is depicted in Table~\ref{table:all-performance}). All the models outperform the existing decompilers. As in the previous case, the first evaluation method has significant differences with the two last ones (which obtain similar results). For the second method, we found that at least 48\% of the real functions must be included in the training dataset to obtain accuracy convergence (Figure~\ref{figure:all-r}). With this percentage of real functions, our models are able to predict functions of programmers whose code is not included in the training set (\textit{i.e.}, there are no significant differences between the second and third evaluation method).

Table~\ref{table:all-performance} shows how gradient boosting obtains the best performance: 0.786 accuracy and 0.791 F\textsubscript{1}-score for the third evaluation method. Comparing these values with the existing decompilers, the performance of gradient boosting is from 96.5\% (accuracy) to 163.6\% (F\textsubscript{1}-score) higher than the decompiler with the highest performance (IDA). Therefore, the gradient boosting vs IDA benefit is increased by 90.1\% when predicting high-level C types.

\begin{table}
\centering
\scriptsize
\begin{tabular}{ll|rrrrrrrrrr}
\toprule
& & \multicolumn{10}{c}{Predicted class} \\
& & \texttt{bool} & \texttt{char} & \texttt{short}  & \texttt{int}  & \texttt{pointer} & \texttt{struct} & \texttt{long long} & \texttt{float} & \texttt{double} & \texttt{void} \\
\midrule
\multirow{10}{*}{\begin{sideways}{Actual class}\end{sideways}}
& \texttt{bool} & \textbf{\textcolor{darkgreen}{491}} & \textbf{\textcolor{darkred}{49}} & 0 & 8 & 10 & 0 & 39 & 0 & 0 & 3 \\
& \texttt{char} & \textbf{\textcolor{darkred}{231}} & \textbf{\textcolor{darkgreen}{212}} & 64 & 20 & 23 & 0 & 45 & 0 & 0 & 5 \\
& \texttt{short} & 6 & 76 & 379 & 45 & 36 & 0 & 49 & 0 & 0 & 9 \\
& \texttt{int} & 6 & 44 & 72 & \textbf{\textcolor{darkgreen}{279}} & \textbf{\textcolor{darkred}{111}} & \textbf{\textcolor{darkred}{21}} & 59 & 0 & 0 & 8 \\
& \texttt{pointer} & 12 & 8 & 0 & \textbf{\textcolor{darkred}{35}} & \textbf{\textcolor{darkgreen}{419}} & \textbf{\textcolor{darkred}{72}} & 47 & 0 & 0 & 7 \\
& \texttt{struct} & 0 & 0 & 0 & \textbf{\textcolor{darkred}{1}} & \textbf{\textcolor{darkred}{12}} & \textbf{\textcolor{darkgreen}{587}} & 0 & 0 & 0 & 0 \\
& \texttt{long long} & 5 & 8 & 0 & 6 & 28 & 4 & 543 & 0 & 1 & 5 \\
& \texttt{float} & 0 & 0 & 0 & 0 & 11 & 0 & 48 & 355 & 184 & 2 \\
& \texttt{double} & 0 & 0 & 0 & 0 & 12 & 0 & 44 & 177 & 365 & 2 \\
& \texttt{void} & 4 & 1 & 1 & 1 & 4 & 0 & 48 & 0 & 0 & 541 \\
\bottomrule
\end{tabular}
\caption{Confusion matrix of the model in Table~\ref{table:confusion-matrix-before}, including the generalizations of Table~\ref{table:advanced-generalization-patterns}.}
\label{table:confusion-matrix-after}
\end{table}

\begin{table}
\centering
\footnotesize
\begin{tabular}{llrrrrrr}
\toprule
& & Accuracy gain & Precision gain & Recall gain & F\textsubscript{1}-score gain \\
\midrule
\multirow{2}{*}{\begin{sideways}{\texttt{INT\_1}}\end{sideways}}
& \texttt{bool} & 0.84\% & 5.85\% & 1.66\% & 3.98\% \\
& \texttt{char} & 1.01\% & 14.86\% & 18.44\% & 17.01\% \\
\midrule
\multirow{3}{*}{\begin{sideways}{\texttt{INT\_4}}\end{sideways}}
& \texttt{int} & 1.81\% & 16.70\% & 51.63\% & 37.76\% \\
& \texttt{pointer} & 0.58\% & 1.69\% & 14.79\% & 7.90\% \\
& \texttt{struct} & 2.56\% & 21.23\% & 0.51\% & 11.55\% \\
\bottomrule
\end{tabular}
\caption{Performance gains obtained for high-level type classification, when the generalizations in Table~\ref{table:advanced-generalization-patterns} are included in the dataset.}
\label{table:performace-gains}
\end{table}

Table~\ref{table:confusion-matrix-after} shows the new values of the confusion matrix presented in Table~\ref{table:confusion-matrix-before}, running the same experiment with the new generalizations. The performance gains obtained when classifying types with the same size and representation are summarized in Table~\ref{table:performace-gains}. We obtain a 10.5\% average F\textsubscript{1}-score gain for 1-byte-size types and 19.1\% for types of 4 bytes, due to the additional generalizations detailed in Table~\ref{table:advanced-generalization-patterns}.

\section{Extracted patterns}
\label{section:extracted-patterns}

In addition to creating models for classifying return types, our dataset can be used to discover and document binary patterns to be included in existing decompilers. That is, we can mine the dataset to document binary patterns associated with high-level return types. That documentation can be helpful to improve the implementation of current decompilers. This is the objective of this section.

We discover binary patterns with association rules that correlate RET and POST CALL patterns with return types of function. Since the dataset has a high number of features, we first select the most important features with the five feature-selection algorithms described in Section~\ref{subsection:feature-selection}. We choose the intersection of the feature sets selected by the classifiers. Then, we run the Apriori algorithm for association rule mining~\cite{Rakesh1994}, saving the rules whose consequent is a return type. 

In this paper, we only analyze the rules with 100\% confidence (\textit{i.e.}, the consequent always holds when the antecedent is true). In this way, the association rules retrieved represent a mechanism to document those RET and POST CALL binary patterns that are unambiguously associated with a high-level return type.

Table~\ref{table:association-rules} shows some of the rules obtained with at most two antecedents. The rest of them are detailed in~\cite{technical-report}. The support of each rule is the relative frequency of instances covered by a rule. Rules with very low support are not included in Table~\ref{table:association-rules}.

The \texttt{cdecl} calling convention~\cite{calling-conventions} returns 32-bit values through \texttt{eax}; \texttt{ax} is used for 16-bit values, and \texttt{al} for 8-bits. 64-bit integers are returned via \texttt{edx} and \texttt{eax} registers. The 32- and 64-bit real values are returned through \texttt{st0}. The main problem is to determine whether such registers are actually returning their values to the caller, or they just hold temporary values of previous computations. Another important problem, as mentioned, is to classify the high-level types with equal size and representation.

\begingroup
\centering
\scriptsize
\begin{longtable}{c|llr}
\toprule
& Antecedents & Consequent & Support \\
\midrule
\endfirsthead
\toprule
& Antecedents & Consequent & Support \\
\midrule
\endhead
\multicolumn{4}{c}{(continues)}\\
\midrule
\endfoot
\bottomrule
\addlinespace
\addlinespace
\caption{Example association rules obtained from the dataset. \textit{reg} variables represent registers, \textit{literal} integer literals, \textit{address} absolute addresses, \textit{*address} absolute addresses dereferences and \textit{offset} relative addresses.}
\label{table:association-rules}
\endlastfoot

\multirow{9}{*}{1} & (RET) & \multirow{9}{*}{\texttt{bool}} & \multirow{9}{*}{0.00059} \\*
& \hspace{4ex}\texttt{mov} \texttt{al}, \textit{literal} & & \\*
& \hspace{4ex}\textit{callee\_epilogue} & & \\*
& & & \\*
& (POST CALL) & & \\*
& \hspace{4ex}\textit{caller\_epilogue} & & \\*
& \hspace{4ex}\texttt{movzx} \texttt{edx}, \texttt{al} & & \\*
& & & \\*
& where: \textit{literal} $\in$ \{\texttt{0}, \texttt{1}\} & & \\

\midrule

\multirow{5}{*}{2} & (RET) & \multirow{5}{*}{\texttt{char}} & \multirow{5}{*}{0.00536} \\*
& \hspace{4ex}\texttt{mov} \texttt{al}, \textit{literal} & & \\*
& \hspace{4ex}\textit{callee\_epilogue} & & \\*
& & & \\*
& where: \textit{literal} $\in$ \{\texttt{0}, \texttt{1}\} & & \\

\midrule

\multirow{12}{*}{3} & (RET) & \multirow{12}{*}{\texttt{char}} & \multirow{12}{*}{0.00207} \\*
& \hspace{4ex}\textit{binary\_op} \texttt{eax} \textit{arg\textsubscript{1}} & & \\*
& \hspace{4ex}\textit{callee\_epilogue} & & \\*
& & & \\*
& (POST CALL) & & \\*
& \hspace{4ex}\textit{caller\_epilogue} & & \\*
& \hspace{4ex}\textit{mov} \textit{arg\textsubscript{2}}, \texttt{al} & & \\*
& & & \\*
& where: \textit{binary\_op} $\in$ \{\texttt{imul}, \texttt{sub}, \texttt{add}, \texttt{sar}, \texttt{sal}, \texttt{shr}, \texttt{shl}, \texttt{xor}, \texttt{or}, \texttt{and}\} & & \\*
& \hspace{8ex}\textit{arg\textsubscript{1}} $\in$ \{\textit{reg}, \texttt{[}\textit{reg}\texttt{]}, \textit{*address}, \textit{literal}\} & & \\*
& \hspace{8ex}\textit{mov} $\in$ \{\texttt{mov}, \texttt{movzx}\} & & \\*
& \hspace{8ex}\textit{arg\textsubscript{2}} $\in$ \{\textit{reg}, \texttt{[}\textit{reg}\texttt{]}, \textit{*address}\} & & \\

\midrule

\multirow{11}{*}{4} & (RET) & \multirow{11}{*}{\texttt{char}} & \multirow{11}{*}{0.00023} \\*
& \hspace{4ex}\texttt{idiv} \textit{arg\textsubscript{1}} & & \\*
& \hspace{4ex}\textit{callee\_epilogue} & & \\*
& & & \\*
& (POST CALL) & & \\*
& \hspace{4ex}\textit{caller\_epilogue} & & \\*
& \hspace{4ex}\textit{mov} \textit{arg\textsubscript{2}}, \texttt{al} & & \\*
& & & \\*
& where: \textit{arg\textsubscript{1}} $\in$ \{\textit{reg}, \texttt{[}\textit{reg}\texttt{]}, \textit{*address}\} & & \\*
& \hspace{8ex}\textit{mov} $\in$ \{\texttt{mov}, \texttt{movzx}\} & & \\*
& \hspace{8ex}\textit{arg\textsubscript{2}} $\in$ \{\textit{reg}, \texttt{[}\textit{reg}\texttt{]}, \textit{*address}\} & & \\

\midrule

\multirow{11}{*}{5} & (RET) & \multirow{11}{*}{\texttt{char}} & \multirow{11}{*}{0.00049} \\*
& \hspace{4ex}\textit{unary\_op} \texttt{eax} & & \\*
& \hspace{4ex}\textit{callee\_epilogue} & & \\*
& & & \\*
& (POST CALL) & & \\*
& \hspace{4ex}\textit{caller\_epilogue} & & \\*
& \hspace{4ex}\textit{mov} \textit{arg}, \texttt{al} & & \\*
& & & \\*
& where: \textit{unary\_op} $\in$ \{\texttt{not}, \texttt{neg}\} & & \\*
& \hspace{8ex}\textit{mov} $\in$ \{\texttt{mov}, \texttt{movzx}\} & & \\*
& \hspace{8ex}\textit{arg} $\in$ \{\textit{reg}, \texttt{[}\textit{reg}\texttt{]}, \textit{*address}\} & & \\

\midrule

\multirow{3}{*}{6} & (POST CALL) & \multirow{3}{*}{\texttt{short}} & \multirow{3}{*}{0.01190} \\*
& \hspace{4ex}\textit{caller\_epilogue} & & \\*
& \hspace{4ex}\texttt{cwde} & & \\

\midrule

\multirow{6}{*}{7} & (POST CALL) & \multirow{6}{*}{\texttt{short}} & \multirow{6}{*}{0.02455} \\*
& \hspace{4ex}\textit{caller\_epilogue} & & \\*
& \hspace{4ex}\textit{mov} \textit{arg}, \texttt{ax} & & \\*
& & & \\*
& where: \textit{mov} $\in$ \{\texttt{mov}, \texttt{movzx}, \texttt{movsx}\} & & \\*
& \hspace{8ex}\textit{arg} $\in$ \{\texttt{eax}, \texttt{ecx}, \texttt{edx}, \texttt{cx}, \texttt{si}, \texttt{[}\textit{reg}\texttt{]}, \textit{*address}\} & & \\

\midrule

\multirow{6}{*}{8} & (RET) & \multirow{6}{*}{\texttt{short}} & \multirow{6}{*}{0.00871} \\*
& \hspace{4ex}\textit{mov} \texttt{ax}, \textit{arg} & & \\*
& \hspace{4ex}\textit{callee\_epilogue} & & \\*
& & & \\*
& where: \textit{mov} $\in$ \{\texttt{mov}, \texttt{movzx}, \texttt{movsx}\} & & \\*
& \hspace{8ex}\textit{arg} $\in$ \{\texttt{dx}, \texttt{al}, \texttt{cx}, \texttt{cl}, \texttt{[}\textit{reg}\texttt{]}, \textit{*address}\} & & \\

\midrule

\multirow{14}{*}{9} & (RET) & \multirow{14}{*}{\texttt{int}} & \multirow{14}{*}{0.01594} \\*
& \hspace{4ex}\textit{cond\_jmp} \textit{offset\textsubscript{1}} & & \\*
& \hspace{4ex}\texttt{mov} \texttt{[}\textit{reg\textsubscript{1}}\texttt{]}, \textit{literal\textsubscript{1}} & & \\*
& \hspace{4ex}\texttt{jmp} \textit{offset\textsubscript{2}} & & \\*
& \hspace{4ex}\texttt{mov} \texttt{[}\textit{reg\textsubscript{1}}\texttt{]}, \textit{literal\textsubscript{2}} & & \\*
& \hspace{4ex}\texttt{mov} \texttt{eax}, \texttt{[}\textit{reg\textsubscript{1}}\texttt{]} & & \\*
& \hspace{4ex}\textit{callee\_epilogue} & & \\*
& & & \\*
& where: \textit{cond\_jmp} $\in$ \{\texttt{jo}, \texttt{jno}, \texttt{js}, \texttt{jns}, \texttt{je}, \texttt{jz}, \texttt{jne}, \texttt{jnz}, \texttt{jb}, \texttt{jnae}, \texttt{jc}, & & \\*
& \hspace{8ex}{\tiny$\hookrightarrow$}\hspace{2ex}\texttt{jnb}, \texttt{jae}, \texttt{jnc}, \texttt{jbe}, \texttt{jna}, \texttt{ja}, \texttt{jnbe}, \texttt{jl}, \texttt{jnge}, \texttt{jge}, \texttt{jnl}, \texttt{jle}, & & \\*
& \hspace{8ex}{\tiny$\hookrightarrow$}\hspace{2ex}\texttt{jng}, \texttt{jg}, \texttt{jnle}, \texttt{jp}, \texttt{jpe}, \texttt{jnp}, \texttt{jpo}, \texttt{jcxz}, \texttt{jecxz}\} & & \\*
& \hspace{8ex}\textit{literal\textsubscript{1}}, \textit{literal\textsubscript{2}} $\in$ \{\texttt{0}, \texttt{1}\} & \\*
& \hspace{8ex}\textit{literal\textsubscript{1}} $\neq$ \textit{literal\textsubscript{2}} & & \\*
& \hspace{8ex}\textit{offset\textsubscript{1}} $\neq$ \textit{offset\textsubscript{2}} & & \\

\midrule

\multirow{6}{*}{10} & (RET) & \multirow{6}{*}{\texttt{int}} & \multirow{6}{*}{0.00103} \\*
& \hspace{4ex}\textit{div} \texttt{ecx} & & \\*
& \hspace{4ex}\texttt{mov} \texttt{eax}, \texttt{edx} & & \\*
& \hspace{4ex}\textit{callee\_epilogue} & & \\*
& & & \\*
& where: \textit{div} $\in$ \{\texttt{div}, \texttt{idiv}\} & & \\

\midrule

\multirow{9}{*}{11} & (RET) & \multirow{9}{*}{\texttt{int}} & \multirow{9}{*}{0.00221} \\*
& \hspace{4ex}\texttt{mov} \texttt{eax}, \textit{literal} & & \\*
& \hspace{4ex}\textit{callee\_epilogue} & & \\*
& & & \\*
& (POST CALL) & & \\*
& \hspace{4ex}\textit{caller\_epilogue} & & \\*
& \hspace{4ex}\texttt{mov} \textit{arg}, \texttt{eax} & & \\*
& & & \\*
& where: \textit{arg} $\in$ \{\texttt{[}\textit{reg}\texttt{]}, \textit{*address}\} & & \\

\midrule

\multirow{5}{*}{12} & (RET) & \multirow{5}{*}{\texttt{pointer}} & \multirow{5}{*}{0.00949} \\*
& \hspace{4ex}\textit{binary\_op} \texttt{eax}, \textit{address} & & \\*
& \hspace{4ex}\textit{callee\_epilogue} & & \\*
& & & \\*
& where: \textit{binary\_op} $\in$ \{\texttt{mov}, \texttt{movzx}, \texttt{movsx}, \texttt{add}, \texttt{sub}\} & & \\

\midrule

\multirow{3}{*}{13} & (RET) & \multirow{3}{*}{\texttt{pointer}} & \multirow{3}{*}{0.01304} \\*
& \hspace{4ex}\texttt{lea} \texttt{eax}, \texttt{[}\textit{reg}\texttt{]} & & \\*
& \hspace{4ex}\textit{callee\_epilogue} & & \\

\midrule

\multirow{3}{*}{14} & (RET) & \multirow{3}{*}{\texttt{struct}} & \multirow{3}{*}{0.06321} \\*
& \hspace{4ex}\texttt{mov} \texttt{eax}, \texttt{[ebp+8]} & & \\*
& \hspace{4ex}\textit{callee\_epilogue} & & \\

\midrule

\multirow{3}{*}{15} & (RET) & \multirow{3}{*}{\texttt{long long}} & \multirow{3}{*}{0.04442} \\*
& \hspace{4ex}\texttt{cdq} & & \\*
& \hspace{4ex}\textit{callee\_epilogue} & & \\

\midrule

\multirow{6}{*}{16} & (RET) & \multirow{6}{*}{\texttt{long long}} & \multirow{6}{*}{0.01063} \\*
& \hspace{4ex}\texttt{mov} \texttt{edx}, \textit{arg} & & \\*
& \hspace{4ex}\textit{callee\_epilogue} & & \\*
& & & \\*
& where: \textit{arg} $\in$ \{\textit{reg}, \texttt{[}\textit{reg}\texttt{]}, \textit{*address}, \textit{literal}\} & & \\

\midrule

\multirow{7}{*}{17} & (RET) & \multirow{7}{*}{\texttt{float}} & \multirow{7}{*}{0.03817} \\*
& \hspace{4ex}\texttt{fstp} \texttt{[}\textit{reg}\texttt{]} & & \\*
& \hspace{4ex}\texttt{fld} \texttt{[}\textit{reg}\texttt{]} & & \\*
& \hspace{4ex}\textit{callee\_epilogue} & & \\*
& & & \\*
& where: opcode(\texttt{fstp})[0] = 0xD9 & & \\
& \hspace{8ex}opcode(\texttt{fld})[0] = 0xD9 & & \\

\midrule

\multirow{6}{*}{18} & (POST CALL) & \multirow{6}{*}{\texttt{float}} & \multirow{6}{*}{0.01545} \\*
& \hspace{4ex}\textit{caller\_epilogue} & & \\*
& \hspace{4ex}\texttt{fstp} \textit{arg} & & \\*
& & & \\*
& where: opcode(\texttt{fstp})[0] = 0xD9 & & \\*
& \hspace{8ex}\textit{arg} $\in$ \{\texttt{[}\textit{reg}\texttt{]}, \textit{*address}\} & & \\

\midrule

\multirow{6}{*}{19} & (RET) & \multirow{6}{*}{\texttt{double}} & \multirow{6}{*}{0.06783} \\*
& \hspace{4ex}\texttt{fld} \textit{arg} & & \\*
& \hspace{4ex}\textit{callee\_epilogue} & & \\*
& & & \\*
& where: opcode(\texttt{fld})[0] = 0xDD & & \\*
& \hspace{8ex}\textit{arg} $\in$ \{\texttt{[}\textit{reg}\texttt{]}, \textit{*address}\} & & \\

\midrule

\multirow{6}{*}{20} & (RET) & \multirow{6}{*}{\texttt{void}} & \multirow{6}{*}{0.00817} \\*
& \hspace{4ex}\texttt{mov} \textit{arg\textsubscript{1}} \textit{arg\textsubscript{2}} & & \\*
& \hspace{4ex}\textit{callee\_epilogue} & & \\*
& & & \\*
& where: \textit{arg\textsubscript{1}} $\in$ \{\texttt{[}\textit{reg}\texttt{]}, \textit{*address}\} & & \\*
& \hspace{8ex}\textit{arg\textsubscript{2}} $\in$ \{\textit{reg}, \texttt{[}\textit{reg}\texttt{]}, \textit{*address}, \textit{literal}\} & & \\

\end{longtable}
\endgroup

Rules 1-5 return the value with \texttt{al}, so they classify \texttt{bool} and \texttt{char} types. The \texttt{ax} register in rules 6-8 is used to return \texttt{short}. Rules 9-14 analyze \texttt{eax} for 4-byte-size types. Rules 15 and 16 check \texttt{edx} to infer \texttt{long long}, and rules 17-19 use \texttt{st0} to return \texttt{float} and \texttt{double}. The last rule checks that the value copied before returning from the function call is not moved to a register (but to a memory address), classifying the function type as \texttt{void}.

Some functions return literal values, such as \texttt{true} or \texttt{32}. The value of those literals is used by some patterns to infer the return type. For example, rule 2 classifies as \texttt{char} the 1-byte type returned when the returned literal is neither \texttt{0} nor \texttt{1} (low-level representation of \texttt{false} and \texttt{true}). The opposite is not true; when \texttt{0} or {1} is returned, it could be a character (\texttt{`}\textbackslash\texttt{0'} character is widely used in C). For this reason, rule 1 adds a POST CALL check after the invocation. If \texttt{0} or \texttt{1} is returned and it is moved to \texttt{edx} with zero extension using \texttt{movzx} (\textit{i.e.}, high bits are set to zero, without sign), the type is \texttt{bool}; for chars, \texttt{movsx} is used instead (copy with sign). Likewise, rule 12 uses address literals to classify pointers.

Our system also detects operations that can only be applied to certain types. For example, division (\texttt{div} and \texttt{idiv}) can be applied to neither \texttt{pointer} nor \texttt{struct}. Therefore, rule 10 infers a 4-byte type to \texttt{int}, when division operations are applied to it. Rule 13 classifies as \texttt{pointer} any 4-byte type where a \texttt{lea} instruction is used, since \texttt{lea} loads a memory address into the target register (\texttt{eax}).

\begin{table}
\centering
\footnotesize
\begin{tabular}{lcc}
\toprule
Textual & \multicolumn{2}{c}{Opcodes} \\
representation & Float & Double \\
\midrule
\texttt{fstp} \textit{reg\textsubscript{fp}} &  \hspace{2ex}\texttt{D9} \texttt{??} \texttt{??} & \hspace{2ex}\texttt{DD} \texttt{??} \texttt{??} \\
\texttt{fld} \textit{reg\textsubscript{fp}} &  \hspace{2ex}\texttt{D9} \texttt{??} \texttt{??} & \hspace{2ex}\texttt{DD} \texttt{??} \texttt{??} \\
\bottomrule
\end{tabular}
\caption{Binary encodings of \texttt{fld} and \texttt{fstp} instructions.}
\label{table:float-point-opcodes}
\end{table}

Another classification mechanism used by our models is based on the binary representation of assembly instructions. For example, the \texttt{fstp} and \texttt{fld} assembly instructions for real numbers share the start of the binary opcode\footnote{Opcode stands for operation code. It is the portion of the numeric representation of an assembly instruction that specifies the operation to be performed.} (Table~\ref{table:float-point-opcodes}). When they operate with 32-bit floating-point numbers, the opcode starts with \texttt{0xD9}. However, their opcode starts with \texttt{0xDD} when applied to 64-bit operands. This difference is used by rules 17-19 to tell the difference between \texttt{float} and \texttt{double}.

The classifiers generated with our dataset also detect binary patterns of the code generation templates implemented by compilers~\cite{Muchnick1998}. For example, rule 9 detects the code generation template used by \texttt{cl} to return the result of a comparison as an \texttt{int}. Of course, these kinds of templates are compiler dependent, so the compiler used should be discovered before using them~\cite{Rosenblum2010}.

Rule 14 is another rule for a particular code generation template. As described in Section~\ref{subsection:grouping-types}, \texttt{cl} performs a code transformation to return \texttt{struct} types (Figure~\ref{figure:code-transformation}). The \texttt{struct} is passed as an argument, and its memory address is actually returned as a \texttt{pointer}. This code transformation generates a particular sequence of assembly instructions that our models use to identify structs among types of 4-bytes size.

Although the assembly instructions used to return a value (RET patterns) are very important to infer return types, the binary code used after the invocation (POST CALL patterns) is also valuable. For example, the usage of \texttt{ax} just after an invocation is used by rules 6 and 7 to identify \texttt{short} types. Another example is rule 18, which stores the returned floating-point value from the mathematical coprocessor stack.

Finally, our system is also able to combine RET and CALL POST patterns to infer return types. Since these kinds of rules are more specific, they commonly have low support and high confidence. For example, the best rule found to classify \texttt{bool} with one RET pattern provides 76\% confidence; whereas rule 1 provides 100\% confidence by adding a CALL POST pattern. These types of rules commonly classify types among others with similar size and representation, such as rules~1, 3 and 4 (1 byte), and rule~11 (4 bytes).

\section{Conclusions}
\label{section:conclusions}

We show how machine learning can be used with a large amount of binary code to improve existing decompilers, particularly for the problem of inferring the high-level type returned by functions. Our system obtains 79.1\% F\textsubscript{1}-score when predicting return types of code written by programmers whose code has not been used in the training set; whereas the best existing decompiler achieves 30\% F\textsubscript{1}-score. Gradient boosting is the best classification algorithm for the given dataset.

Additionally, we discuss and document the binary patterns found to classify return types. The classification rules combine binary patterns for returning expressions and the opcodes after function invocation. They focus on how data are passed between the function and the caller. The binary patterns found not only distinguish among different sizes and representations of data, but also among types with the same binary size and representation. The publication and documentation of these patterns~\cite{technical-report} can be used for different purposes, including the improvement of current decompilers.

We plan to study the appropriateness of sequence classifiers such as recurrent neural networks (RNN) for inferring the return type of functions. Its ability to exploit the order of binary instructions, for both the return and invocation sequences, seems to be adequate for the return type problem~\cite{Chua2017}. Another possible direction to improve our work is the augmentation of pattern generalizations with information gathered from symbolic execution and type-based constraints~\cite{Mycroft1999}. Apart from improving the performance of the classifiers, those techniques could also be used to give more information about the C composite types. We also plan to use other compilers, compilation parameters and architectures to check if the proposed method is applicable to those scenarios. Finally, we would like to apply this methodology to reconstruct the type of other language constructs such as global and local variables, and function parameters.

The binaries and source code of our system, the C code corpus used to create the dataset, the Python code to build and evaluate the models, the hyper-parameters selected for each model, the datasets, and the evaluation data used in this article can be freely downloaded from \newline \url{http://www.reflection.uniovi.es/bigcode/download/2021/expertsa}

\section*{Acknowledgments}
\label{section:acknowledgments}

This work has been partially funded by the Spanish Department of Science, Innovation and Universities: project RTI2018-099235-B-I00. The authors have also received funds from the University of Oviedo through its support to official research groups (GR-2011-0040).

\bibliographystyle{elsarticle-num}
\bibliography{bibliography}

\begin{thebibliography}{10}
\expandafter\ifx\csname url\endcsname\relax
  \def\url#1{\texttt{#1}}\fi
\expandafter\ifx\csname urlprefix\endcsname\relax\def\urlprefix{URL }\fi
\expandafter\ifx\csname href\endcsname\relax
  \def\href#1#2{#2} \def\path#1{#1}\fi

\bibitem{Horspool1980}
R.~N. Horspool, N.~Marovac, {An Approach to the Problem of Detranslation of
  Computer Programs}, The Computer Journal 23~(3) (1980) 223--229.

\bibitem{Cifuentes1994}
C.~Cifuentes, Reverse compilation techniques, Ph.D. thesis, School of Computing
  Science, Queensland University of Technology, AU (1994).

\bibitem{VanEmmerik2007}
M.~J. Van~Emmerik, Static single assignment for decompilation, Ph.D. thesis,
  School of Information Technology and Electrical Engineering, University of
  Queensland, AU (2007).

\bibitem{Ortin2016}
F.~Ortin, J.~Escalada, O.~Rodriguez-Prieto, {Big Code: New Opportunities for
  Improving Software Construction}, Journal of Software 11~(11) (2016)
  1083--1088.

\bibitem{Raychev2015}
V.~Raychev, M.~Vechev, A.~Krause, {Predicting Program Properties from "Big
  Code"}, in: Proceedings of the 42nd Annual ACM SIGPLAN-SIGACT Symposium on
  Principles of Programming Languages - POPL '15, Vol.~50, ACM Press, New York,
  New York, USA, 2015, pp. 111--124.

\bibitem{Karaivanov2014}
S.~Karaivanov, V.~Raychev, M.~Vechev, {Phrase-Based Statistical Translation of
  Programming Languages}, in: Proceedings of the 2014 ACM International
  Symposium on New Ideas, New Paradigms, and Reflections on Programming {\&}
  Software - Onward! '14, ACM Press, New York, New York, USA, 2014, pp.
  173--184.

\bibitem{Yamaguchi2014}
F.~Yamaguchi, N.~Golde, D.~Arp, K.~Rieck, {Modeling and Discovering
  Vulnerabilities with Code Property Graphs}, in: 2014 IEEE Symposium on
  Security and Privacy, IEEE, 2014, pp. 590--604.

\bibitem{Ortin2020}
F.~Ortin, O.~Rodriguez-Prieto, N.~Pascual, M.~Garcia, Heterogeneous tree
  structure classification to label java programmers according to their
  expertise level, Future Generation Computer Systems 105 (2020) 380 -- 394.

\bibitem{Chua2017}
Z.~L. Chua, S.~Shen, P.~Saxena, Z.~Liang, {Neural Nets Can Learn Function Type
  Signatures From Binaries}, in: 26th USENIX Security Symposium (USENIX
  Security 17), {USENIX Association}, Vancouver, BC, 2017, pp. 99--116.

\bibitem{He2018}
J.~He, P.~Ivanov, P.~Tsankov, V.~Raychev, M.~Vechev, {Debin: Predicting Debug
  Information in Stripped Binaries}, in: Proceedings of the 2018 ACM SIGSAC
  Conference on Computer and Communications Security, ACM, New York, NY, USA,
  2018, pp. 1667--1680.

\bibitem{KatzDeborah2018}
D.~S. Katz, J.~Ruchti, E.~Schulte, {Using recurrent neural networks for
  decompilation}, in: 25th IEEE International Conference on Software Analysis,
  Evolution and Reengineering, SANER 2018 - Proceedings, Vol. 2018-March,
  Institute of Electrical and Electronics Engineers Inc., 2018, pp. 346--356.

\bibitem{Schulte2018}
E.~Schulte, J.~Ruchti, M.~Noonan, D.~Ciarletta, A.~Loginov, {Evolving Exact
  Decompilation}, in: Proceedings 2018 Workshop on Binary Analysis Research,
  Internet Society, Reston, VA, 2018, pp. 1--11.

\bibitem{Rumelhart1986}
D.~E. Rumelhart, G.~E. Hinton, R.~J. Williams, {Learning representations by
  back-propagating errors}, Nature 323~(6088) (1986) 533--536.

\bibitem{Bengio2003}
Y.~Bengio, R.~Ducharme, P.~Vincent, C.~Janvin, A neural probabilistic language
  model, J. Mach. Learn. Res. 3~(null) (2003) 1137–1155.

\bibitem{Lafferty2001}
J.~D. Lafferty, A.~McCallum, F.~C.~N. Pereira, {Conditional Random Fields:
  Probabilistic Models for Segmenting and Labeling Sequence Data}, in: In
  Proceedings of the 18th International Conference on Machine Learning, ICML
  '01, Morgan Kaufmann Publishers Inc., San Francisco, CA, USA, 2001, pp.
  282--289.

\bibitem{Brumley2011}
D.~Brumley, I.~Jager, T.~Avgerinos, E.~J. Schwartz, {BAP}: A binary analysis
  platform, in: G.~Gopalakrishnan, S.~Qadeer (Eds.), Computer Aided
  Verification, Springer Berlin Heidelberg, Berlin, Heidelberg, 2011, pp.
  463--469.

\bibitem{KatzOmer2019}
O.~Katz, Y.~Olshaker, Y.~Goldberg, E.~Yahav, {Towards Neural Decompilation},
  ArXiv (2019).
\newblock \href {http://arxiv.org/abs/1905.08325} {\path{arXiv:1905.08325}}.

\bibitem{Fu2019}
C.~Fu, H.~Chen, H.~Liu, X.~Chen, Y.~Tian, F.~Koushanfar, Zhao, {Coda: An
  End-to-End Neural Program Decompiler}, in: H.~Wallach, H.~Larochelle,
  A.~Beygelzimer, F.~d\textquotesingle Alch{\'{e}}-Buc, E.~Fox, R.~Garnett
  (Eds.), Advances in Neural Information Processing Systems 32, Curran
  Associates, Inc., 2019, pp. 3708--3719.

\bibitem{Kalchbrenner2013}
N.~Kalchbrenner, P.~Blunsom, Recurrent continuous translation models, in:
  Proceedings of the 2013 Conference on Empirical Methods in Natural Language
  Processing, Association for Computational Linguistics, Seattle, Washington,
  USA, 2013, pp. 1700--1709.

\bibitem{Sutskever2014}
I.~Sutskever, O.~Vinyals, Q.~V. Le, {Sequence to Sequence Learning with Neural
  Networks}, Advances in Neural Information Processing Systems 4~(January)
  (2014) 3104--3112.
\newblock \href {http://arxiv.org/abs/1409.3215} {\path{arXiv:1409.3215}}.

\bibitem{Henderson2014}
M.~Henderson, B.~Thomson, S.~J. Young, Robust dialog state tracking using
  delexicalised recurrent neural networks and unsupervised adaptation, 2014
  IEEE Spoken Language Technology Workshop (SLT) (2014) 360--365.

\bibitem{Mycroft1999}
A.~Mycroft, Type-based decompilation (or program reconstruction via type
  reconstruction), in: Proceedings of the 8th European Symposium on Programming
  Languages and Systems, ESOP ’99, Springer-Verlag, Berlin, Heidelberg, 1999,
  p. 208–223.

\bibitem{Milner1978}
R.~Milner, A theory of type polymorphism in programming, Journal of Computer
  and System Sciences 17 (1978) 348--375.

\bibitem{Caballero2016}
J.~Caballero, Z.~Lin, {Type Inference on Executables}, ACM Computing Surveys
  48~(4) (2016) 1--35.

\bibitem{Xue2019}
H.~Xue, S.~Sun, G.~Venkataramani, T.~Lan, {Machine Learning-Based Analysis of
  Program Binaries: A Comprehensive Study}, IEEE Access 7 (2019) 65889--65912.

\bibitem{Rosenblum2008}
N.~N. Rosenblum, X.~Zhu, B.~B. Miller, K.~Hunt, {Learning to analyze binary
  computer code}, in: Proceedings of the 23rd Conference on Artificial
  Intelligence, Chicago, 2008, pp. 798--804.

\bibitem{Bao2014}
T.~Bao, J.~Burket, M.~Woo, R.~Turner, D.~Brumley, {BYTEWEIGHT: Learning to
  recognize functions in binary code}, in: Proceedings of the 23rd USENIX
  Security Symposium, {USENIX} Association, San Diego, CA, 2014, pp. 845--860.

\bibitem{Briandais1959}
R.~De~La~Briandais, File searching using variable length keys, in: Western
  Joint Computer Conference, IRE-AIEE-ACM ’59 (Western), Association for
  Computing Machinery, New York, NY, USA, 1959, p. 295–298.

\bibitem{Shin2015}
E.~C.~R. Shin, D.~Song, R.~Moazzezi, {Recognizing Functions in Binaries with
  Neural Networks}, in: 24th USENIX Security Symposium, {USENIX} Association,
  Washington, D.C., 2015, pp. 611--626.

\bibitem{Rosenblum2010}
N.~E. Rosenblum, B.~P. Miller, X.~Zhu, {Extracting compiler provenance from
  program binaries}, in: Proceedings of the 9th ACM SIGPLAN-SIGSOFT workshop on
  Program analysis for software tools and engineering, PASTE '10, ACM Press,
  Toronto, Ontario, Canada, 2010, pp. 21--28.

\bibitem{Rosenblum2011}
N.~Rosenblum, B.~P. Miller, X.~Zhu, {Recovering the toolchain provenance of
  binary code}, in: Proceedings of the 2011 International Symposium on Software
  Testing and Analysis, ISSTA '11, ACM Press, Toronto, Ontario, Canada, 2011,
  pp. 100--110.

\bibitem{Ucci2017}
D.~Ucci, L.~Aniello, R.~Baldoni, {Survey of Machine Learning Techniques for
  Malware Analysis}, Computers {\&} Security 81 (2017) 123--147.
\newblock \href {http://arxiv.org/abs/1710.08189} {\path{arXiv:1710.08189}}.

\bibitem{Alazab2010}
M.~Alazab, S.~Venkatraman, P.~Watters, M.~Alazab, Zero-day malware detection
  based on supervised learning algorithms of api call signatures, in:
  Proceedings of the Ninth Australasian Data Mining Conference - Volume 121,
  AusDM ’11, Australian Computer Society, Inc., AUS, 2011, p. 171–182.

\bibitem{Rathore2018}
H.~Rathore, S.~Agarwal, S.~K. Sahay, M.~Sewak, Malware detection using machine
  learning and deep learning, Lecture Notes in Computer Science (2018)
  402–411.

\bibitem{Escalada2017}
J.~Escalada, F.~Ortin, T.~Scully, {An Efficient Platform for the Automatic
  Extraction of Patterns in Native Code}, Scientific Programming 2017 (2017)
  1--16.

\bibitem{cnerator}
F.~Ortin, J.~Escalada, {Cnerator: A C source code generator},
  https://github.com/computationalreflection/cnerator (2021).

\bibitem{scikit-learn}
F.~Pedregosa, G.~Varoquaux, A.~Gramfort, V.~Michel, B.~Thirion, O.~Grisel,
  M.~Blondel, P.~Prettenhofer, R.~Weiss, V.~Dubourg, J.~Vanderplas, A.~Passos,
  D.~Cournapeau, M.~Brucher, M.~Perrot, E.~Duchesnay, Scikit-learn: Machine
  learning in {P}ython, Journal of Machine Learning Research 12 (2011)
  2825--2830.

\bibitem{Kohavi1997}
R.~Kohavi, G.~H. John, {Wrappers for feature subset selection}, Artificial
  Intelligence 97~(1-2) (1997) 273--324.

\bibitem{paperwebpage}
J.~Escalada, F.~Ortin, Improving type information inferred by decompilers with
  supervised machine learning (support material webpage),
  http://www.reflection.uniovi.es/bigcode/download/2021/expertsa (2021).

\bibitem{hex-rays-decompiler}
{Hex-Rays}, {Hex Rays Decompiler}, https://www.hex-rays.com/products/decompiler
  (2021).

\bibitem{hex-rays-ida}
{Hex-Rays}, {Hex Rays IDA}, https://www.hex-rays.com/products/ida (2021).

\bibitem{Guilfanov2001}
I.~Guilfanov, {Simple type system for program reengineering}, in: Proceedings
  Eighth Working Conference on Reverse Engineering, IEEE Computer Society,
  2001, pp. 357--361.

\bibitem{Guilfanov2008}
I.~Guilfanov, {Decompilers and beyond}, in: Black Hat USA, 2008, pp. 1--12.

\bibitem{retdec}
P.~Matula, {RetDec}, https://github.com/avast/retdec (2021).

\bibitem{Kroustek2015}
J.~K\v{r}oustek, Retargetable analysis of machine code, Ph.D. thesis, Faculty
  of Information Technology, Brno University of Technology, CZ (2015).

\bibitem{snowman}
Y.~Danilov, {Snowman}, https://github.com/yegord/snowman (2021).

\bibitem{Troshina2009}
K.~Troshina, A.~Chernov, Y.~Derevenets, {C Decompilation: Is It Possible?}, in:
  M.~A. Bulyonkov, R.~Gl{\"{u}}ck (Eds.), Proceedings of International Workshop
  on Program Understanding, Ershov Institute of Informatics Systems, Siberian
  Branch of the Russian Academy of Sciences, Altai Mountains, Russia, 2009, pp.
  18--27.

\bibitem{Fokin2011}
A.~Fokin, E.~Derevenetc, A.~Chernov, K.~Troshina, {SmartDec: Approaching C++
  Decompilation}, in: 2011 18th Working Conference on Reverse Engineering,
  IEEE, 2011, pp. 347--356.

\bibitem{hopper}
{Cryptic Apps EURL}, {Hopper, the macOS and Linux disassembler},
  https://www.hopperapp.com (2021).

\bibitem{disc}
S.~Kumar, {DisC - Decompiler for TurboC},
  https://www.debugmode.com/dcompile/disc.htm (2021).

\bibitem{rec}
G.~Caprino, {REC Decompiler}, http://www.backerstreet.com/rec/rec.htm (2021).

\bibitem{Schwartz2013}
E.~J. Schwartz, M.~Woo, D.~Brumley, J.~Lee, {Native x86 Decompilation Using
  Semantics-Preserving Structural Analysis and Iterative Control-Flow
  Structuring}, in: USENIX Security Symposium, USENIX, Washington, D.C., 2013,
  pp. 353--368.

\bibitem{Yakdan2015}
K.~Yakdan, S.~Eschweiler, E.~Gerhards-Padilla, M.~Smith, {No More Gotos:
  Decompilation Using Pattern-Independent Control-Flow Structuring and
  Semantics-Preserving Transformations}, in: Proceedings 2015 Network and
  Distributed System Security Symposium, Internet Society, Reston, VA, 2015,
  pp. 1--15.

\bibitem{Yakdan2016}
K.~Yakdan, S.~Dechand, E.~Gerhards-Padilla, M.~Smith, {Helping Johnny to
  Analyze Malware: A Usability-Optimized Decompiler and Malware Analysis User
  Study}, in: Proceedings - 2016 IEEE Symposium on Security and Privacy,
  Institute of Electrical and Electronics Engineers Inc., 2016, pp. 158--177.

\bibitem{Georges2007}
A.~Georges, D.~Buytaert, L.~Eeckhout, Statistically rigorous {J}ava performance
  evaluation, SIGPLAN Not. 42~(10) (2007) 57–76.

\bibitem{Yang1999}
Y.~Yang, X.~Liu, A re-examination of text categorization methods, in:
  Proceedings of the 22nd Annual International ACM SIGIR Conference on Research
  and Development in Information Retrieval, SIGIR ’99, Association for
  Computing Machinery, New York, NY, USA, 1999, p. 42–49.

\bibitem{Sokolova2009}
M.~Sokolova, G.~Lapalme, A systematic analysis of performance measures for
  classification tasks, Information Processing \& Management 45~(4) (2009) 427
  -- 437.

\bibitem{Opitz2019}
J.~Opitz, S.~Burst, Macro {F}1 and macro {F}1, ArXiv (2019).
\newblock \href {http://arxiv.org/abs/1911.03347} {\path{arXiv:1911.03347}}.

\bibitem{KernighanRitchie1978}
B.~W. Kernighan, D.~M. Ritchie, The C Programming Language, 2nd Edition,
  Prentice Hall Professional Technical Reference, 1988.

\bibitem{Rakesh1994}
R.~Agrawal, R.~Srikant, Fast algorithms for mining association rules in large
  databases, in: Proceedings of the 20th International Conference on Very Large
  Data Bases, VLDB ’94, Morgan Kaufmann Publishers Inc., San Francisco, CA,
  USA, 1994, p. 487–499.

\bibitem{technical-report}
J.~Escalada, F.~Ortin, Association rules obtained from the dataset described in
  \textit{Improving type information inferred by decompilers with supervised
  machine learning},
  http://www.reflection.uniovi.es/bigcode/download/2021/expertsa/tr.pdf (2021).

\bibitem{calling-conventions}
{Microsoft}, {Calling Conventions | Microsoft Docs},
  https://docs.microsoft.com/en-us/cpp/cpp/calling-conventions (2021).

\bibitem{Muchnick1998}
S.~S. Muchnick, Advanced Compiler Design and Implementation, Morgan Kaufmann
  Publishers Inc., San Francisco, CA, USA, 1998.

\end{thebibliography}

\end{document}